%% file: Book_Chapter.tex
\documentclass{article}

\usepackage{amsmath}
\usepackage{amssymb}
\usepackage{algorithm}
\usepackage{amsthm}
\usepackage{graphicx}
\usepackage{subfig}
\usepackage{enumerate}
\usepackage{color}
\usepackage[colorlinks]{hyperref}
\usepackage{cite}
\usepackage{commath}
\usepackage{multirow}
\usepackage{algorithmic}

\definecolor{darkred}{RGB}{150,0,0}
\definecolor{darkgreen}{RGB}{0,150,0}
\definecolor{darkblue}{RGB}{0,0,200}
\hypersetup{colorlinks=true, linkcolor=darkred, citecolor=darkgreen, urlcolor=darkblue}

\newcommand{\aaa}{\mathbf{a}}

\newcommand{\eee}{\mathbf{e}}
\newcommand{\x}{\mathbf{x}}
\newcommand{\y}{\mathbf{y}}
\newcommand{\z}{\mathbf{z}}
\newcommand{\f}{\mathbf{f}}

\newcommand{\w}{\mathbf{w}}

\newcommand{\D}{\mathbf{D}}
\newcommand{\F}{\mathbf{F}}

\newcommand{\II}{\mathbf{I}}

\newcommand{\Y}{\mathbf{Y}}
\newcommand{\Z}{\mathbf{Z}}

\newcommand{\W}{\mathbf{W}}
\newcommand{\X}{\mathbf{X}}

\newcommand{\for}{\quad \textrm{for} \quad}
\newcommand{\aand}{\quad \textrm{and} \quad}
\newcommand{\trace}{ \textrm{trace} }
\newcommand{\otherwise}{ \textrm{otherwise} }

\newcommand{\eps}{\epsilon}

\newcommand{\edit}{}

\begin{document}

%%%%%%%%%%%%%%%%%%%%%%%%%%%%%%%%%%%%%%%%%%%%%%%%%%

\title{Phase Retrieval: An Overview of Recent Developments}
\author{
\begin{tabular}[t]{c@{\extracolsep{5em}}c@{\extracolsep{5em}}c} 
Kishore Jaganathan$^\dagger$ & Yonina C. Eldar$^\ddagger$  & Babak Hassibi$^\dagger$
\end{tabular}
\\ \\
$^\dagger$Department of Electrical Engineering, Caltech \\
$^\ddagger$Department of Electrical Engineering, Technion, Israel Institute of Technology     
}
\date{}

\maketitle

%%%%%%%%%%%%%%%%%%%%%%%%%%%%%%%%%%%%%%%%%%%%%%%%%%

%\begin{abstract}

%The problem of phase retrieval is a classic one in optics and arises when one is interested in recovering an unknown signal from the magnitude (intensity) of its Fourier transform. While there have existed quite a few approaches to phase retrieval, recent developments in compressed sensing and convex optimization-based signal recovery have inspired a host of new ones. This work presents an overview of these approaches. 

%Since phase retrieval, by its very nature, is ill-posed, to make the problem meaningful one needs to either assume prior structure on the signal (e.g., sparsity) or obtain additional measurements (e.g., masks, structured illuminations). For both the cases, we review conditions for the identifiability of the signal, as well as practical algorithms for signal recovery. In particular, we demonstrate that it is possible to robustly and efficiently identify an unknown signal solely from phaseless Fourier measurements, a fact with potentially far-reaching implications.
%\end{abstract}

%%%%%%%%%%%%%%%%%%%%%%%%%%%%%%%%%%%%%%%%%%%%%%%%%%

\input{Introduction.tex}

\input{SPR_Introduction.tex}

\input{SPR_Uniqueness.tex}

\input{SPR_Algorithms.tex}

\input{SPR_Simulations.tex}

%%%%%%%%%%%%%%%%%%%%%%%%%%%%%%%%%%%%%%%%%%%%%%%%%%

\input{PRM_Introduction.tex}

\input{PRM_Uniqueness_Algorithms.tex}

\input{PRM_Simulations.tex}

%%%%%%%%%%%%%%%%%%%%%%%%%%%%%%%%%%%%%%%%%%%%%%%%%%

\input{STFTPR_Introduction.tex}

\input{STFTPR_Uniqueness.tex}

\input{STFTPR_Algorithms.tex}

\input{STFTPR_Simulations.tex}

\section{Conclusions}

In this chapter, we reviewed some of the recent progress on phase retrieval. In many cases, we demonstrated that an unknown signal can be robustly and efficiently recovered from a set of phaseless Fourier measurements. In particular, we first considered the problem of sparse phase retrieval. We noted that most sparse signals can be uniquely identified from their autocorrelation, and presented two efficient and robust algorithms (TSPR and GESPAR). Then, we considered the problem of phase retrieval using masks. We presented various masks which allow for unique and robust recovery of the unknown signal. Finally, we treated the problem of STFT phase retrieval. We noted that most signals can be uniquely identified from their STFT magnitude, and suggested an efficient and robust algorithm (STliFT) for signal recovery. These results clearly have many practical applications in optics and imaging systems where measuring the phase is often challenging. They may also have ramifications to other areas of engineering, such as radar, wireless communications, and more, where the possibility of avoiding having to measure the phase could result in simpler and cost-effective practical systems.

{\bf Acknowledgement}: The authors would like to thank Prof. Mordechai Segev and Oren Cohen for introducing them to the STFT phase retrieval problem, and for many insightful discussions regarding phase retrieval and optics.

%%%%%%%%%%%%%%%%%%%%%%%%%%%%%%%%%%%%%%%%%%%%%%%%%%

%%%%%%%%%%%%%%%%%%%%%%%%%%%%%%%%%%%%%%%%%%%%%%%%%%

\end{document}

%% file: Introduction.tex
\section{Introduction}

In many physical measurement systems, one can only measure the power spectral density, i.e., the magnitude-square of the Fourier transform of the underlying signal.  For example, in an optical setting, detection devices like {\edit CCD} cameras and photosensitive films cannot measure the phase of a light wave and instead measure the photon flux.
In addition, at a large enough distance from the imaging plane the field is given by the Fourier transform of the image (up to a known phase factor).
Thus, in the far field, optical devices essentially measure the Fourier transform magnitude. Since the phase encodes a lot of the structural content of the image, important information is lost. The problem of reconstructing a signal from its Fourier magnitude is known as phase retrieval \cite{patt1, patt2}. This reconstruction problem is one with a rich history and arises in many areas of engineering and applied physics, including optics \cite{walther}, X-ray crystallography \cite{millane}, astronomical imaging \cite{dainty}, speech processing \cite{rabiner}, computational biology \cite{stef}, blind deconvolution \cite{baykal} and more.

\begin{figure}
\begin{center}
\includegraphics[scale=0.7]{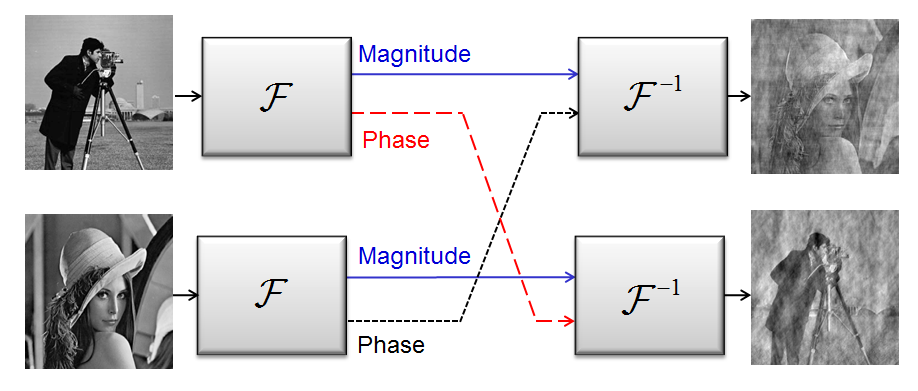}  
\end{center}
\caption{A synthetic example demonstrating the importance of Fourier phase in reconstructing a signal from its Fourier transform (courtesy of \cite{eldarmagazine}).}
\label{fig:PhaseImportant}
\end{figure}

Reconstructing a signal from its Fourier magnitude alone is generally a very difficult task. It is well known that Fourier  phase is quite often more important than Fourier  magnitude in reconstructing a signal from its Fourier transform \cite{oppenheim1}. To demonstrate this fact, a synthetic example, courtesy of \cite{eldarmagazine}, is provided in Figure \ref{fig:PhaseImportant}. The figure shows the result of the following numerical simulation: Two images are Fourier transformed, their Fourier phases are swapped and then they are inverse Fourier transformed. The result clearly demonstrates the importance of Fourier phase. Therefore, simply ignoring the phase and performing an inverse Fourier transform does not lead to satisfactory recovery. Instead,  algorithmic phase retrieval can be used, offering a means for recovering
the phase from the given magnitude measurements and possibly additional prior knowledge, providing an alternative to sophisticated
measurement setups as in holography which attempt to directly measure the phase by requiring interference with another known field.

To set up the phase retrieval problem mathematically, we focus on the discretized one-dimensional (1D) setting. Let $\x = ( x[0], x[1], \ldots , x[N-1] )^T$ be a signal of length $N$ such that it has non-zero values only within the interval $[0,N-1]$. Denote by $\y = ( y[0], y[1], \ldots , y[N-1] )^T$ its $N$ point discrete Fourier transform (DFT) and let $\z=( z[0], z[1] , \ldots, z[N-1] )^T$ be the Fourier magnitude-square measurements $z[m] = \abs{y[m]}^2$. Phase retrieval can be mathematically stated as:
\begin{align}
\label{PRf}
&\textrm{find} \hspace{2cm} \x  \\
\nonumber & \textrm{subject to} \hspace{1cm} z[m] = \abs{\left<\f_m,\x\right>}^2 \for 0 \leq m \leq N-1,
\end{align}
{\edit where $\f_m$ is the conjugate of the $m$th column of the $N$ point DFT matrix, with elements $e^{i2\pi \frac{mn}{N}}$, and $\left<.,.\right>$ is the inner product operator.}

\subsection{Classic Approaches}

For any given Fourier magnitude, the Fourier phase can be chosen from an $N$-dimensional set. Since distinct phases correspond to different signals in general, the feasible set of (\ref{PRf}) is an $N$-dimensional manifold, rendering phase retrieval a very ill-posed problem. One approach to try and overcome the ill-posedness is to employ oversampling by using an $M>N$ point DFT (see \cite{fienup0} for a comprehensive survey). A typical choice is $M=2N$. The term {\em oversampling} is used in this chapter to refer to measurements from the $M=2N$ point DFT.

Phase retrieval with oversampling can be equivalently stated as the problem of reconstructing a signal from its autocorrelation measurements $\aaa = ( a[0], a[1], \ldots , a[N-1] )^T$, i.e.,
\begin{align}
\label{PRal}
& \textrm{find} \hspace{2cm}  {\x} \\
\nonumber & \textrm{subject to}  \hspace{1cm} a[m] = \sum_{j=0}^{N-1-m} x[j] x^\star[m+j] \for 0 \leq m \leq N-1.
\end{align}
This is because the length $M=2N$ DFT of $\aaa$ is given by $\z$.

Observe that the operations of time-shift, conjugate-flip and global phase-change on the signal do not affect the autocorrelation, because of which there are trivial ambiguities. Signals obtained by these operations are considered equivalent, and in most applications it is good enough if any equivalent signal is recovered. For example, in astronomy, where the underlying signal corresponds to stars in the sky, or in X-ray crystallography, where the underlying signal corresponds to atoms or molecules in a crystal, equivalent solutions are equally informative \cite{dainty, millane}.

In the 1D setup, it has been shown, using spectral factorization,  that there is no uniqueness and the feasible
set of (\ref{PRal}) can include up to $2^N$ non-equivalent solutions \cite{hofstetter}. While this is a significant improvement (when compared to the feasible set of (\ref{PRf})), $2^N$ is still a prohibitive number because of which phase retrieval with oversampling remains ill-posed.  Furthermore, adding support constraints on $\x$ does not help to ensure uniqueness. However, for higher dimensions (2D and above), it has been shown by Hayes \cite{hayes} using dimension counting that, with the exception of a set of signals of measure zero, phase retrieval with oversampling is well posed up to trivial ambiguities.

As we discuss further below, to guarantee unique identification in the 1D case it is necessary to assume
(or enforce) additional constraints on the unknown signal such as sparsity or to introduce specific redundancy into the measurements.
Even when unique identification of the underlying signal is theoretically possible, it is unclear how to find the unique solution efficiently and robustly. Earlier approaches to phase retrieval were based on alternating projections, pioneered by the work of Gerchberg and Saxton \cite{gerchberg}. In this framework, phase retrieval is reformulated as the following least-squares problem:
\begin{equation}
\label{PRleastsquares}
\min_\x \quad \sum_{m=0}^{M-1} \left( z[m] - \abs{\left<\f_m,\x\right>}^2 \right)^2.
\end{equation}
The Gerchberg-Saxton (GS) algorithm attempts to minimize this non-convex objective by starting with a random initialization and iteratively imposing the time domain (support) and Fourier magnitude constraints using projections. The details of the various steps are provided in Algorithm \ref{alg:GS}. The objective is shown to be monotonically decreasing as the iterations progress. However, since the projections are between a convex set (for the time domain constraints) and a non-convex set (for the Fourier magnitude constraints), the convergence is often to a local minimum, due to which the algorithm has limited recovery abilities even in the noiseless setting.

\begin{algorithm}
\caption{Gerchberg-Saxton  (GS) Algorithm \label{alg:GS}}
\begin{algorithmic}
\STATE {\bf Input}: Fourier magnitude-square measurements $\z$
\STATE {\bf Output:} Estimate $\hat{\x}$ of the underlying signal
\STATE {\bf Initialize}: Choose a random input signal $\x^{(0)}$, $\ell = 0$
\WHILE{halting criterion false}
\STATE $\ell \leftarrow \ell+1$
\STATE Compute the DFT of $\x^{(\ell-1)}$: $\y^{(\ell)}=\F \x^{(\ell-1)}$
\STATE Impose Fourier magnitude constraints: $y'^{(\ell)}[m]=\frac{y^{(\ell)}[m]}{ \abs{y^{(\ell)}[m]} }\sqrt{z[m]}$
\STATE Compute the inverse DFT of $\y'^{(\ell)}$: $\x'^{(\ell)} = {\F}^{-1}\y'^{(\ell)}$
\STATE Impose time-domain constraints to obtain $\x^{(\ell)}$
\ENDWHILE
\STATE return $\hat{\x} \leftarrow {\x}^{(\ell)}$
\end{algorithmic}
\end{algorithm}

Fienup, in his seminal work \cite{fienup}, extended the alternating projection framework by adding additional time-domain restrictions and introducing several variants to the time-domain projection step. The most popular of these is the Hybrid Input-Output (HIO) algorithm. While the basic framework of the HIO technique is similar to the GS method, the former includes an additional time-domain {\em correction step} in order to improve convergence (see \cite{fienup} for details). The HIO algorithm is not guaranteed to converge, and when it does converge, it may be to a local minimum.
Nonetheless, HIO methods and their variants are often used in optical phase retrieval. We refer the readers to \cite{bauschke} and \cite{marchesini} for a theoretical and numerical investigation of these techniques.

\subsection{Recent Approaches}

{\edit Recently, phase retrieval has benefited from a surge of research in both the optics and the mathematical communities due to various new imaging techniques in optics, and advances in modern optimization tools and structured-based information processing \cite{yoninabook1, yoninabook2, yoninabook3}. For example, one imaging modality which has gained large interest in the past $15$ years is coherent diffractive imaging (CDI) \cite{miao} in which an object is illuminated with a
coherent wave and the far-field diffraction intensity pattern is measured.  Phase retrieval algorithms are a key component in enabling CDI imaging.
On the algorithmic side, as we detail further below, ideas of semidefinite relaxation and sparsity have played a key role in modern phase retrieval.

Recent approaches to phase retrieval can be broadly classified into two categories:

(i) {\em Additional prior information}: Inspired by advances in the field of compressed sensing \cite{candesl0, candesmc, yoninabook2, yoninabook3}, various researchers have explored the idea of sparsity as a prior information on the signal \cite{baraniukcpr, vetterli, mukherjee, candespr, eldarcvx, kishoreorig, kishorej, eldargespar, ranieri, kishore2}. A signal of length $N$ is said to be $k$-sparse if it has $k$ locations with non-zero values and $k \ll N$. 
The exact locations and values of
the nonzero elements are not known a priori.
The approach to sparse phase retrieval has been to develop conditions under which only one sparse signal satisfies the autocorrelation measurements and to suggest algorithms which exploit the sparsity to improve convergence and robustness to noise.

(ii) {\em Additional magnitude-only measurements}: Technological advances have enabled the possibility of obtaining more information about the signal via additional magnitude-only measurements. This can be done in various ways depending on the application; common approaches include the use of masks \cite{pfeiffer}, optical gratings \cite{popov}, oblique illuminations \cite{faridian}, and short-time Fourier transform (STFT) magnitude measurements which utilize overlap between adjacent short-time sections \cite{trebino, rodenburg, yang, humphry, kishorestft}. An important line of research is to identify the necessary additional magnitude-only measurements needed in order to render the recovery unique, efficient and robust.

Another popular trend in the phase retrieval literature for analysis purposes is to replace the Fourier measurements with random measurements so that $z[m] = \abs{\left<\mathbf{v}_m,\x\right>}^2$ for random vectors $\mathbf{v}_m$.  This allows to derive uniqueness and recovery guarantees more easily than for Fourier measurements \cite{candespl, ohlsson, li, samet, balan1, balan2, afonso3, yoninapr, sujay}.
Since this review focuses on Fourier phase retrieval which naturally arises in optics, astronomy and more, we do not pursue this line of work here.}

On the algorithmic front, one of the recent popular approaches to treat phase retrieval problems of both categories above is to use semidefinite programming (SDP) methods. SDP algorithms have been shown to yield robust solutions to various quadratic-constrained optimization problems (see \cite{goemans, daspremont, candespl} and references therein). Since phase retrieval results in quadratic constraints, it is natural to use SDP techniques to try and solve such problems \cite{candespr, eldarcvx, kishoreorig}. An SDP formulation of phase retrieval (\ref{PRf}) can be obtained by a procedure popularly known as {\em lifting}--We embed $\x$ in a higher dimensional space using the transformation $\X = \x\x^\star$. The Fourier magnitude measurements are then linear in the matrix $\X$:
\begin{equation}
\nonumber z[m] = \abs{\left<\f_m,\x\right>}^2 = \x^\star \f_m \f_m^\star \x = {\trace}({\f_m\f_m^\star \x \x^\star}) = {\trace}(\f_m\f_m^\star \X).
\end{equation}
Consequently, phase retrieval reduces to finding a rank one positive semidefinite matrix $\X$ which satisfies these affine measurement constraints, leading to the following reformulation:
\begin{align}
 &\textrm{minimize} \hspace{1.2cm}  \textrm{rank}(\X) \nonumber \\
& \nonumber \textrm{subject to} \hspace{1cm}  z[m] = {\trace}(\f_m\f_m^\star \X) \for 0 \leq m \leq N-1\\
& \nonumber \hspace{2.55cm} \X \succcurlyeq 0.
\end{align}
Unfortunately, $\mbox{rank}(\X)$ is a non-convex function of $\X$. To obtain a convex program one possibility is to replace it by the convex surrogate $\mbox{trace}(\X)$ \cite{fazel1, candesmc}, resulting in the convex SDP
\begin{align}
\label{SPRL}
 &\textrm{minimize} \hspace{1.2cm}  \trace(\X)  \\
& \nonumber \textrm{subject to} \hspace{1cm}  z[m] = {\trace}(\f_m\f_m^\star \X) \for 0 \leq m \leq N-1\\
& \nonumber \hspace{2.55cm} \X \succcurlyeq 0.
\end{align}
This approach is referred to as PhaseLift \cite{candespr}.

{\edit SDP algorithms are known to be robust in general. However, due to the high dimensional transformation involved, they can be computationally demanding. A recent alternative is to use gradient-based techniques with appropriate initialization. For the random measurement case, an alternating minimization algorithm is proposed in \cite{sujay}, and a non-convex algorithm based on Wirtinger Flow is suggested in \cite{wirtinger}. Beyond these general purpose techniques, efficient phase retrieval algorithms have been developed which take advantage of sparsity \cite{beck} and specific magnitude-only measurements. In particular, clever mask designs allow for recovery using simple combinatorial algorithms \cite{afonso2, kishoreprm, kannan2}.}

The purpose of this chapter is to provide an overview of some of the recent developments, both theoretical and algorithmic, in phase retrieval. In particular, we focus on sparse phase retrieval, the use of masks, and recovery from the STFT magnitude. We refer the readers to the recent overview \cite{eldarmagazine} which contains applications to optics as well as a more comprehensive review of the history of phase retrieval and some of the algorithmic aspects.

The rest of the chapter is organized as follows. In Section 2, we motivate sparse phase retrieval, which is the problem of reconstructing a sparse signal from its Fourier magnitude. Section 3 considers phase retrieval using masks.  In Section 4, we study STFT phase retrieval in which the measurements correspond to the STFT magnitude. For each of these three problems, we provide a literature survey of the available uniqueness guarantees and describe various recovery algorithms. Section 5 concludes the chapter.

%% file: SPR_Introduction.tex
\section{Sparse Phase Retrieval}

\label{sec:spr}

In many applications of phase retrieval, the underlying signals are naturally sparse due to the physical nature of the setup. For instance, electron microscopy deals with sparsely distributed atoms or molecules \cite{millane}, while astronomical imaging tends to consider sparsely distributed stars \cite{dainty}.  More generally, a sparsity prior corresponds to having advance knowledge that the unknown signal has some characteristic structure, or, equivalently, that it has a small number of degrees of freedom. The simplest setting is when the basis in which the object is represented compactly is known in advance. This basis is referred to as the sparsity basis, or dictionary. When such a dictionary is not given a priori, it can often be learned from the measurements themselves \cite{blindcs} or from data with similar features that may be available from other sources \cite{yoninabook2}. Sparsity priors have been used extensively in many fields of engineering and statistics and are known to model well various classes of images and signals. {\edit Recently, the use of sparsity has also become popular in optical applications including holography \cite{fresnel},  super-resolution and sub-wavelength imaging \cite{yoav, yoninanature, gazit, yoninanature2, superres, subwave} and ankylography \cite{mutzafi, ankyl}}.

If it is known a priori that the signal of interest is sparse, then one could potentially solve for the sparsest solution satisfying the Fourier magnitude measurements. Our problem can then be written as
\begin{align}
\label{SPR0}
& \textrm{minimize} \hspace{1.1cm}  {\| \x \|_0} \\
\nonumber & \textrm{subject to}  \hspace{1cm} z[m] = \abs{\left<\f_m,\x\right>}^2 \for 0 \leq m \leq M-1,
\end{align}
where $\|.\|_0$ is the $\ell_0$ norm which counts the number of nonzero entries of its argument. When $M=2N$, sparse phase retrieval is equivalent to the problem of reconstructing a sparse signal from its autocorrelation.

%% file: SPR_Uniqueness.tex
\subsection{Uniqueness}

We begin by reviewing existing uniqueness results for sparse phase retrieval (\ref{SPR0}). These results are summarized in Table I. Since the operations of time-shift, conjugate-flip and global phase-change on the signal do not affect its sparsity, similar to standard phase retrieval, recovery is possible only up to these trivial ambiguities.

A general uniqueness result is derived in \cite{kishoreorig}, where it is shown, using dimension counting, that most sparse signals with {\em aperiodic support} can be uniquely identified from their autocorrelation. A signal is said to have periodic, or aperiodic support, if the locations of its non-zero components are uniformly spaced, or not uniformly spaced, respectively. As an example, consider the signal $\x = ( x[0], x[1], x[2], x[3], x[4] )^T$ of length $N=5$.
\begin{enumerate}[(i)]
\item Possible  aperiodic supports: $\{n | x[n] \neq 0\} = \{0 , 1 , 3\}$, $\{1, 2, 4\}$.
\item Possible periodic supports:  $\{n | x[n] \neq 0\} = \{0, 2, 4\}$, $\{0, 1, 2, 3, 4\}$.
\end{enumerate}
Consequently, if the signal of interest is known to have aperiodic support, then sparse phase retrieval is {\em almost surely} well-posed. Note, however, that this still does not provide an efficient robust method for finding the sparse input.
An explanation as to why recovery of a sparse signal with periodic support is generally not possible is provided in \cite{vetterli}. Specifically, it is argued that a sparse signal with periodic support can be viewed as an upsampled version of a non-sparse signal. Sparse phase retrieval in this case is then seen to be equivalent to phase retrieval, because of which most such signals cannot be uniquely identified from their autocorrelation.

In \cite{ranieri}, it is shown that the knowledge of the autocorrelation is sufficient to uniquely identify 1D sparse signals if the autocorrelation is  {\em collision free}, as long as the sparsity $k \neq 6$. A signal $\x$ is said to have a collision free autocorrelation if for all indices  $\{i_1,i_2,i_3,i_4\}$ such that $\{x[i_1], x[i_2], x[i_3], x[i_4]\} \neq 0$, we have $\abs{i_1-i_2} \neq \abs{i_3 -i_4}$. In words, a signal is said to have a collision free autocorrelation if no two pairs of locations with non-zero values in the signal are separated by the same distance. For higher dimensions, the authors  show that the requirement $k \neq 6$ is not necessary. This result has been further refined in \cite{ohlsson2}, where it is shown that $k^2 - k + 1$ Fourier magnitude measurements are sufficient to recover the autocorrelation.

\begin{figure}
\begin{center}
\includegraphics[scale=0.59]{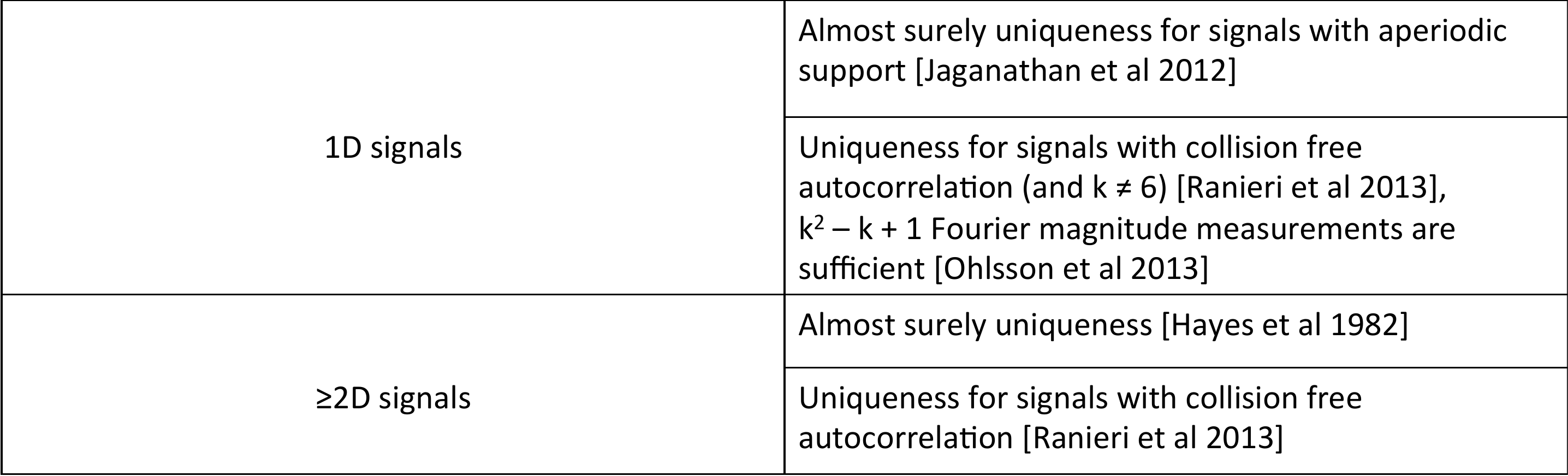}
\end{center}
\begin{center}
{\edit Table I: Uniqueness results for sparse phase retrieval (with $M \geq 2N$ point DFT).}
\end{center}
\end{figure}

%% file: SPR_Algorithms.tex
\subsection{Algorithms}

Phase retrieval algorithms based on alternating projections and SDP have been adapted to solve sparse phase retrieval. In this subsection, we first describe these two adaptations, explain their limitations and then describe two powerful sparse phase retrieval algorithms: TSPR \cite{kishorej} and GESPAR {\cite{eldargespar}. TSPR can {\em provably} recover most $O(N^{1/2-\eps})$-sparse signals up to the trivial ambiguities. Further, for most $O(N^{1/4-\eps})$-sparse signals, the recovery is robust in the presence of noise. {\edit GESPAR has been shown to yield fast and accurate recovery results, and has been used in several phase retrieval optics applications \cite{waveguide, yoninanature2, yoninanature, mutzafi, yoav}}.

{\em (i) Alternating projections}: The Fienup HIO algorithm has been extended to solve sparse phase retrieval by adapting the step involving time domain constraints to promote sparsity. This can be achieved in several ways. For example, the locations with absolute values less than a particular threshold may be set to zero. Alternatively, the $k$ locations with the highest absolute values can be retained and the rest set to zero \cite{mukherjee}. In the noiseless setting, the sparsity constraint partially alleviates the convergence issues if multiple random initializations are considered and the underlying signals are sufficiently sparse. However, in the noisy setting, convergence issues still remain.

{\em (ii) SDP-based methods}: It is well-known that $\ell_1$-minimization, which is a convex surrogate for $\ell_0$-minimization, promotes sparse solutions \cite{candesl0}. Hence, a natural convex program to solve sparse phase retrieval is:
\begin{align}
\label{SPRR}
&\textrm{minimize} \hspace{1.05cm}  \trace(\X) +  \lambda \| \X \|_1 \\
& \nonumber \textrm{subject to} \hspace{0.9cm}  z[m] = \trace(\f_m\f_m^\star \X)   \for  0 \leq m \leq M-1\\
& \nonumber \hspace{2.4cm} \X \succcurlyeq 0,
\nonumber \end{align}
for some regularizer $\lambda > 0$. While this approach has enjoyed considerable success in related problems like generalized phase retrieval for sparse signals \cite{li, ohlsson, samet}, it often fails to solve sparse phase retrieval.

This does not come as a surprise as the issue of trivial ambiguities (due to time-shift and conjugate-flip) is still unresolved. If $\X_{0}=\x_0\x^\star_0$ is the desired sparse solution, then $\tilde{\X}_0=\tilde{\x}_0\tilde{\x}^\star_0$, where $\tilde{\x}_0$ is the conjugate-flipped version of $\x_0$, $\X_j=\x_j\x_j^\star$, where $\x_j$ is the signal obtained by time-shifting $\x_0$ by $j$ units, and $\tilde{\X}_j=\tilde{\x}_j\tilde{\x}_j^\star$, where $\tilde{\x}_j$ is the signal obtained by time-shifting $\tilde{\x}_0$ by $j$ units are also feasible with the same objective value as $\X_0$. Since (\ref{SPRR}) is a convex program, any convex combination of these solutions is feasible as well and has an objective value less than or equal to that of $\X_0$, because of which the optimizer is neither sparse nor rank one. Many iterative heuristics have been proposed to break this symmetry \cite{eldarcvx, candespr, kishore2}. While these heuristics enjoy empirical success, they are computationally expensive due to the fact that each iteration involves the solution of a high dimensional convex program and the number of iterations required for convergence can be high.

\input{TSPR.tex}

\input{GESPAR.tex}

%% file: TSPR.tex
{\em (iii) TSPR}: The convex program (\ref{SPRR}) fails to solve sparse phase retrieval primarily because of issues due to trivial ambiguities, which stem from the fact that the support of the signal is unknown. In order to overcome this, Two-stage Sparse Phase Retrieval (TSPR) is proposed in \cite{kishorej}, which involves: (i) estimating the support of the signal using an efficient algorithm and (ii) estimating the signal values {\em with known support} using the convex program (\ref{SPRR}).

The first stage of TSPR, which involves recovery of the support of the signal (denoted by $V$) from the support of the autocorrelation (denoted by $B$), is equivalent to that of recovering an integer set from its pairwise distances (also known as the {\em Turnpike problem} \cite{skiena, dakic, kishoreturn}). For example, consider the integer set $V=\{2,5,13,31,44\}$. Its pairwise distance set is given by $B=\{0,3,8,11,13,18,26,29,31,39,42\}$. The Turnpike problem is to reconstruct $V$ from  $B$. In \cite{kishorej}, it is argued that due to trivial ambiguities, without loss of generality, one can always construct a solution $U = \{u_0, u_1, \ldots , u_{k-1}\}$ which is a subset of $B$. TSPR, in essence, eliminates all the integers in $B$ that do not belong to $U$ using two instances of the {\em Intersection step} and one instance of the {\em Graph step}. These steps are explained in \cite{kishorej}.
As to the second step, even though is involves lifting, the dimension of the problem is reduced from $N$ to $k$ due to  knowledge of the support from the first stage. In the noisy setting, TSPR considers the pairwise distance set of $B$, along with a series of {\em Generalized intersection steps}, to ensure stable recovery. TSPR provably recovers most $O(N^{1/2-\eps})$-sparse signals efficiently and most $O(N^{1/4-\eps})$-signals robustly. A brief overview of the steps involved is provided in Algorithm \ref{alg:TSPR}. We refer the interested readers to \cite{kishorej} for further details.

%In the noisy setup, a modified version of  is proposed in \cite{kishorej} to make the recovery stable.

\begin{algorithm}
\caption{TSPR (see \cite{kishorej} for noisy setting)}
\label{alg:TSPR}
\textbf{Input:} Autocorrelation measurements $\aaa$ \\
\textbf{Output:} Sparse estimate $\hat{\x}$ of the underlying sparse signal
\begin{itemize}
\item[1)] Obtain $B = \{ ~n~ | ~a[n] \neq 0 ~ \}$
\item[2)] Infer $\{ u_0, u_1, u_{k-1} \}$ from $B$
\item[3)] Intersection step using $u_{1}$: obtain $Z = 0 \cup \left( B \cap (B+u_{1}) \right)$
\item[4)] Graph step using $(Z, B)$: obtain $\{u_{2}, u_3, \ldots , u_t\}$ for $t=\sqrt[3]{\log(k)}$
\item[5)] Intersection step using $\{u_2, u_3, \ldots , u_t\}$: obtain $U$
\item[6)] Obtain $\hat{\X}$ by solving:
\begin{align}
\label{SPRS}
 &\textrm{minimize} \hspace{1cm} \trace(\X) \\
\nonumber & \textrm{subject to} \hspace{0.9cm}   z[m] = \trace(\f_m\f_m^\star \X)   \for  0 \leq m \leq M-1\\
\nonumber & \hspace{2.4cm} X[n_1,n_2] = 0 \quad \textrm{if} \quad  \{n_1,n_2\} \notin U\\
\nonumber & \hspace{2.4cm} \X \succcurlyeq 0
\end{align}
\item[7)] Return $\hat{\x}$, where $\hat{\x} \hat{\x}^\star$ is the best rank one approximation of $\hat{\X}$
\end{itemize}
\end{algorithm}

%% file: GESPAR.tex
{\em (iv) GESPAR}:  In \cite{eldargespar}, a sparse optimization-based greedy search method called GESPAR (GrEedy Sparse PhAse Retrieval) is proposed. Sparse phase retrieval is reformulated as the following sparsity-constrained least-squares problem:
\begin{align}
\label{gesparls}
& \min_{\x} \hspace{1.5cm} \sum_{m=0}^{M-1} \left( z[m]-\abs{\left< \f_m ,  \x \right> }^2 \right)^2 \\
\nonumber & \textrm{subject to} \hspace{0.7cm} \| \x \|_0 \leq k.
\end{align}
GESPAR is a local search method, based on iteratively updating the signal support, and seeking a vector that corresponds to the measurements under the current support. A location-search method is repeatedly invoked, beginning with an initial random support set. Then, at each iteration, a swap is performed between a support and a non-support index. Only two elements are changed in the swap (one in the support and one in the non-support), following the so-called 2-opt method \cite{papa}. Given the support of the signal, phase retrieval is then treated as a non-convex optimization problem, and approximated using the damped Gauss-Newton method \cite{bertsekas}.

{\edit GESPAR has been used in several phase retrieval optics applications, including CDI of 1D objects \cite{yoninanature2, yoninanature}, efficient CDI of sparsely temporally varying objects \cite{yoav} and phase retrieval via waveguide arrays \cite{waveguide}}. A brief overview of the steps involved is provided in Algorithm \ref{algo:gespar}. We refer the interested readers to \cite{eldargespar} for further details.

\begin{algorithm}
\caption{\label{algo:gespar}GESPAR (see \cite{eldargespar} for details)}

\textbf{Input:} Autocorrelation measurements $\aaa$, parameters $\tau$ and ${\rm ITER}$\\
\textbf{Output:} Sparse estimate $\hat{\x}$ of the underlying sparse signal

\textbf{Initialization}: Set $T=0$ and $j=0$
\begin{itemize}
\item[1)] Generate a random support set $S^{(0)}$ of size $k$
\item[2)] Invoke the damped Gauss Newton method with support  $S^{(0)}$, and obtain $\x^{(0)}$
\end{itemize}
\textbf{General Step ($j=1,2,\ldots$)}:
\begin{itemize}
\item[3)] Update support: Let $p$ be the index from $S^{(j-1)}$ corresponding to the component of $\x^{(j-1)}$ with the smallest absolute value and $q$ be the index from the complement of $S^{(j-1)}$ corresponding to the component of $\nabla f(\x^{(j-1)})$ with the highest absolute value, where $\nabla f(\x)$ is the gradient of the least-squares objective function (\ref{gesparls}). Increase $T$ by 1, and make a swap between the indices $p$ and $q$, i.e., ${S'}=(S^{(j-1)}\backslash\{p\})\cup\{q\}$

\item[4)] Minimize with given support: Invoke the damped Gauss Newton method with support $S'$, and obtain ${\x'}$ \\ \\
 If $f({\x'})<f(\x^{(j-1)})$, then set $S^{(j)}={S'},\x^{(j)}={\x'}$,
advance $j$ and go to Step 3. If none of the swaps resulted with a better objective function value, go to Step 1

\end{itemize}
\textbf{Until} $f(\x)<\tau$ or $T>{\rm ITER}$.

Return $\hat{\x} \leftarrow \x^{(j)}$
\end{algorithm}

%% file: SPR_Simulations.tex
\subsection{Numerical Simulations}

In this subsection, we demonstrate the performance of TSPR and GESPAR using numerical simulations.

\begin{figure}
\begin{center}
\subfloat[{Probability of successful recovery for $N = 6400$ and varying choices of  $k$.}]{\includegraphics[scale=0.5]{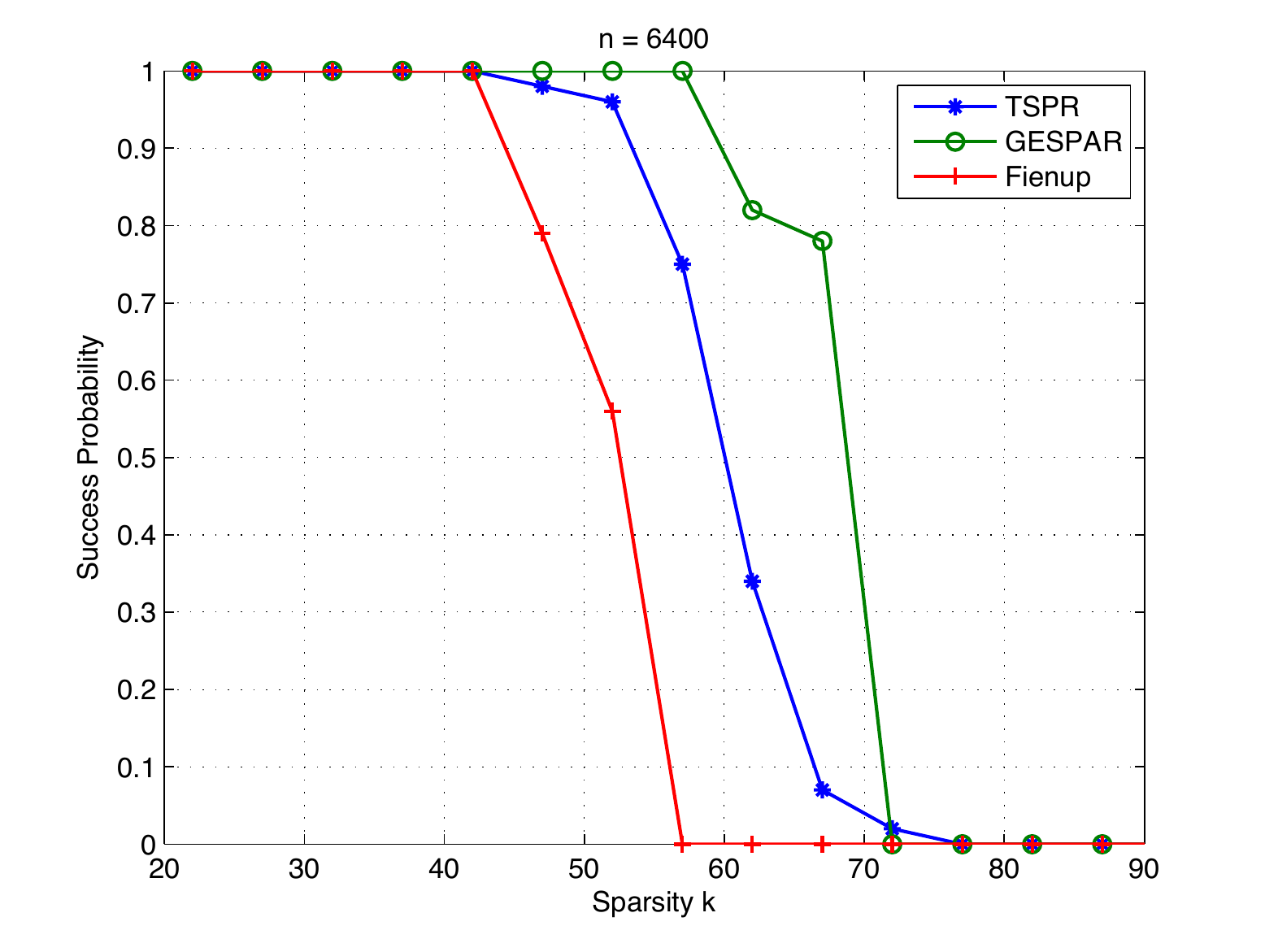} \label{spr_sim3}} \hspace{1cm}
\subfloat[{Probability of successful recovery of TSPR for varying choices of $N$ and $k$.}]{\includegraphics[scale=0.5]{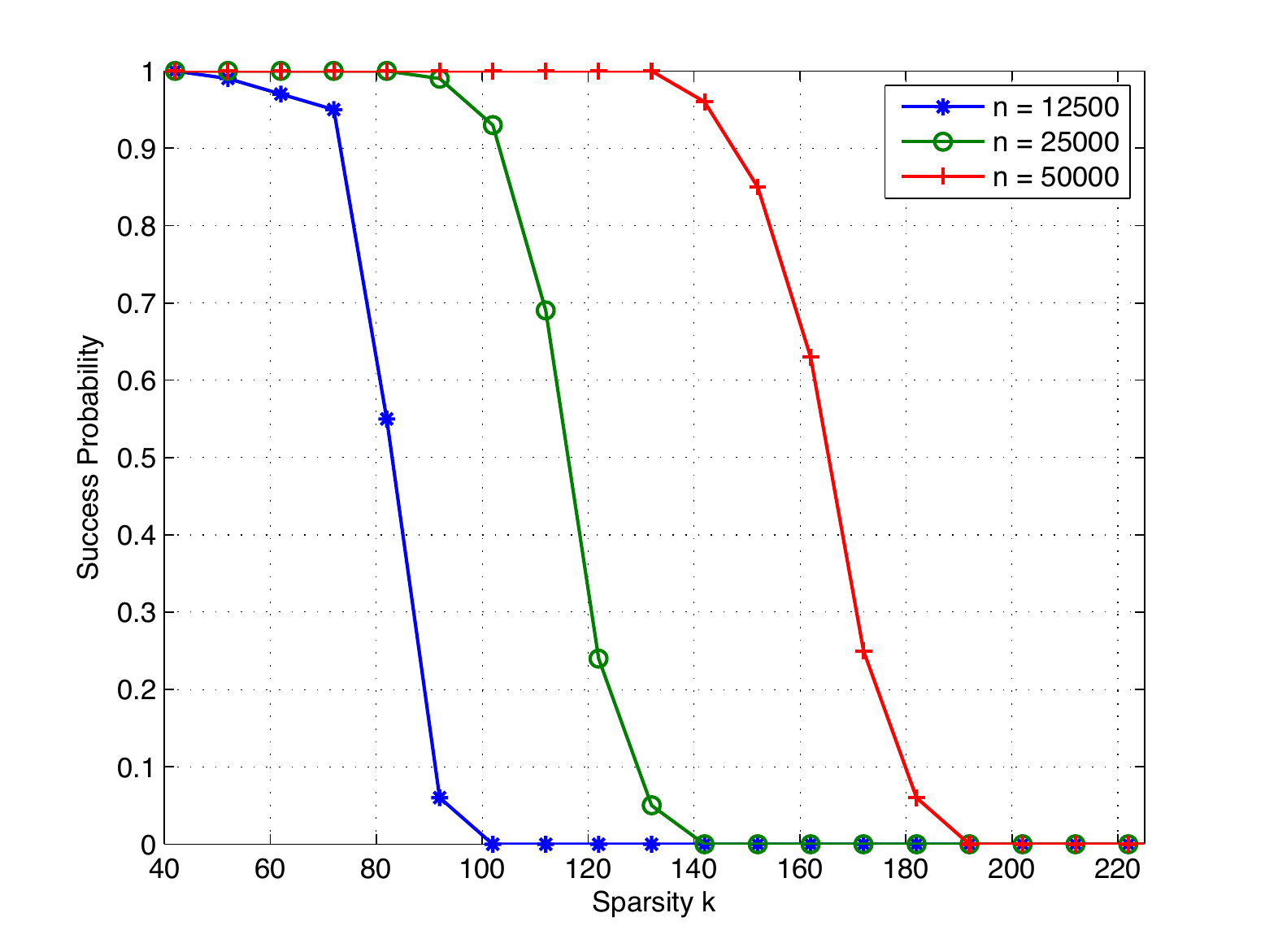}\label{spr_sim1}}
\end{center}
\caption{Performance of various sparse phase retrieval algorithms {\edit (courtesy of \cite{kishorej} and \cite{eldargespar})}.}
\end{figure}

We compare the recovery ability of the Fienup HIO algorithm, TSPR and GESPAR  in the noiseless setting. We consider sparse signals of length $N=6400$ for varying sparsities $20 \leq k \leq 90$. {\edit For each sparsity, $100$ trials are performed. The locations with non-zero values are chosen uniformly at random and the values in the non-zero locations are chosen from an i.i.d. standard normal distribution.} Fienup HIO algorithm is run with $200$ random initializations, GESPAR is run with parameters $\tau = 10^{-4}$ and $ITER = 10000$. The probability of successful recovery of these algorithms is plotted versus sparsity in Figure \ref{spr_sim3}. It can be seen that GESPAR has the best empirical performance among the three algorithms: GESPAR recovers signals with sparsities up to $57$ while TSPR recovers signals with sparsities up to $53$. Both GESPAR and TSPR significantly outperform the Fienup HIO algorithm.

In Figure \ref{spr_sim1}, the probability of successful recovery of TSPR for $N = \{12500, 25000, 50000\}$ is plotted versus sparsity. The $O(N^{1/2-\eps})$ theoretical guarantee of TSPR is clearly verified empirically. For example, the choices $N = 12500$, $k = 80$ and $N = 50000$, $k=160$ have a success probability of $0.5$.

%% file: PRM_Introduction.tex
\section{Phase Retrieval using Masks}

\begin{figure}[t]
\begin{center}
\includegraphics[scale=0.3]{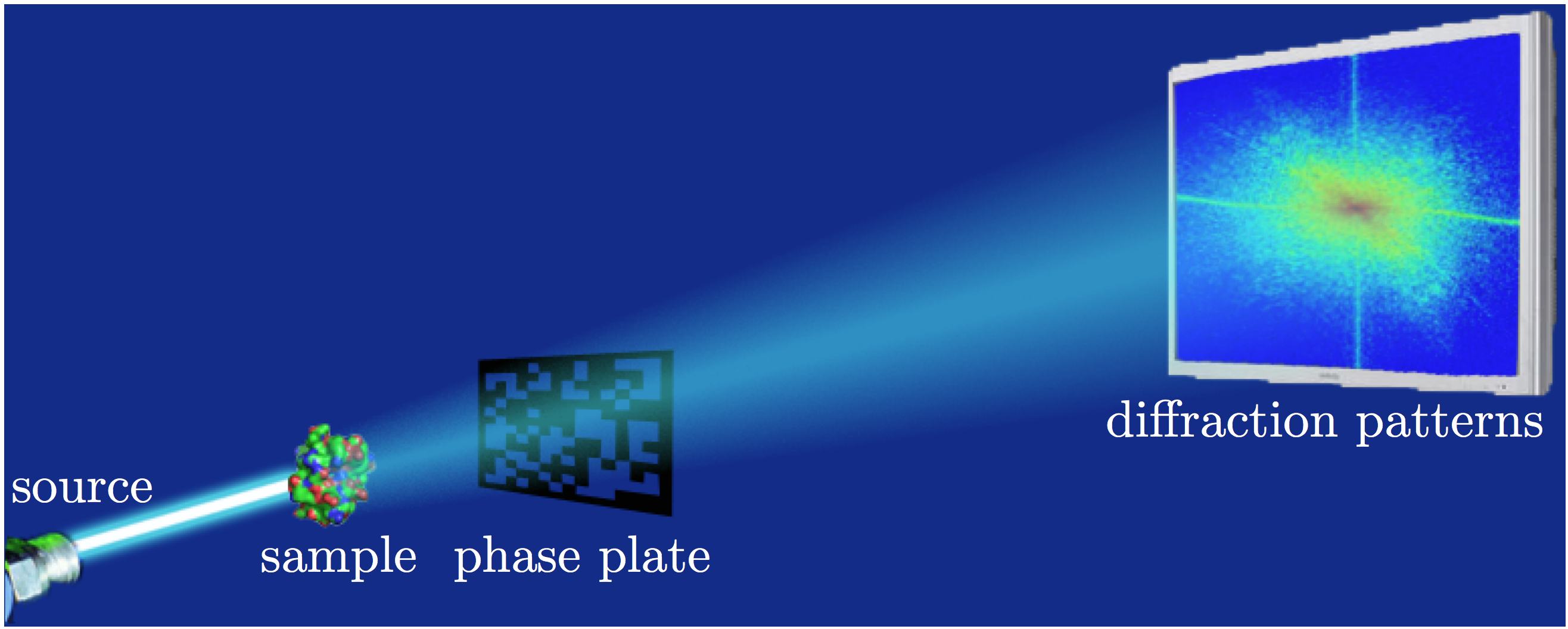}
\end{center}
\caption{A typical setup for phase retrieval using masks or modulated illuminations (courtesy of \cite{mahdi}).}
\label{fig:dif1}
\end{figure}

The use of masked magnitude-only measurements in order to resolve the Fourier phase uniquely has been explored by various researchers. 
The key idea, in a nutshell, is to obtain additional information about the signal using multiple masks in order to mitigate the uniqueness and algorithmic issues of phase retrieval.
There are many ways in which this can be performed in practice, depending on the application. Several such methods are summarized in \cite{candespr}: {\em (i) Masking}: The phase front after the sample is modified by the use of a mask or a phase plate \cite{pfeiffer, liu}. A schematic representation, courtesy of \cite{mahdi}, is provided in Figure \ref{fig:dif1}. {\em (ii) Optical grating}:  The illuminating beam is modulated by the use of optical gratings \cite{popov}, using a setup similar to Figure \ref{fig:dif1}. {\em (iii) Oblique illuminations}: The illuminating beam is modulated to hit the sample at specific angles \cite{faridian}.

Suppose Fourier magnitude-square measurements are collected using $R$ masks (or modulated illuminations). For $0 \leq r \leq R-1$, let $\D_r$ be an $N \times N$ diagonal matrix, corresponding to the $r${th} mask or modulated illumination, with diagonal entries $(d_r[0], d_{r}[1], \ldots , d_{r}[N-1] )$. Let $\Z$ denote the $N \times R$ magnitude-square measurements, such that the $r${th} column of $\Z$ corresponds to the magnitude-square of the $N$ point DFT of the masked signal $\D_r \x$. Phase retrieval using masks then reduces to the following reconstruction problem:
\begin{align}
\label{MPRf}
&\textrm{find} \hspace{2cm} \x  \\
\nonumber & \textrm{subject to} \hspace{1cm} Z[m,r] = \abs{\left<\f_m,\D_r\x\right>}^2 \for 0 \leq m \leq N - 1\aand 0 \leq r \leq R-1.
\end{align}
 
A natural question to ask is how many masks are needed in order to guarantee uniqueness. While this question, in its full generality, is still open, results are available when it comes to uniquely identifying most signals. The best known result is due to \cite{kishoreprm}, where it is shown that two generic masks are sufficient to uniquely identify most signals up to global phase if an $M \geq 2N$ point DFT is considered. While this is a strong identifiability result, it is unclear how to efficiently and robustly recover the underlying signal from these measurements. Another natural question to ask, hence, is how many and which masks allow for unique, efficient and robust recovery of the underlying signal.

%% file: PRM_Uniqueness_Algorithms.tex
\subsection{Uniqueness and Algorithms}

Phase retrieval algorithms based on SDP and stochastic gradient descent (Wirtinger Flow algorithm \cite{wirtinger}) have been adapted to solve phase retrieval using masks uniquely, efficiently and robustly for some choices of masks. Combinatorial algorithms have also been developed for specific mask designs which allow for unique and efficient reconstruction in the noiseless setting. In the following, we survey the main algorithms proposed along with the corresponding choice of masks.

{\em (i) SDP methods}: SDP-based phase retrieval has been adapted to account for masks in \cite{mahdi, kishoreprm} by solving
\begin{align}
\label{LMPRf}
&\textrm{minimize} \hspace{1cm}  \trace(\X) \\
& \nonumber \textrm{subject to} \hspace{0.8cm} Z[m,r] = \trace( ~ {\D_r^\star\f_m\f_m^\star \D_r\X} ~ ) \for 0 \leq m \leq N - 1\aand 0 \leq r \leq R-1\\
& \nonumber \hspace{2.4cm} \X \succcurlyeq 0.
\end{align}
In order to provide recovery guarantees, the masks in \cite{mahdi} are chosen from a random model. In particular, the diagonal matrices $\D_r$ are assumed to be i.i.d. copies of a matrix $\D$, whose entries consist of i.i.d. copies of a random variable $d$ satisfying the following properties:
\begin{equation}
\label{distreq}
\nonumber \mathbb{E}[d] = 0 \quad \quad \mathbb{E}[d^2] = 0 \quad \quad \mathbb{E}|d|^4 = 2 \mathbb{E}|d|^2.
\end{equation}
An example of an admissible random variable is  given by $d = b_1b_2$, where $b_1$ and $b_2$ are independent and distributed as
\begin{equation}
\label{maskex}
b_1 = \begin{cases}
1 \quad &\textrm{with prob.} \quad \frac{1}{4} \\
-1 \quad  &\textrm{with prob.} \quad \frac{1}{4} \\
{i}  \quad & \textrm{with prob.} \quad \frac{1}{4} \\
-{i}  \quad &\textrm{with prob.} \quad \frac{1}{4}
\end{cases}
\quad \quad 
b_2 = \begin{cases}
1 \quad &\textrm{with prob.} \quad \frac{4}{5} \\
\sqrt{6} \quad &\textrm{with prob.} \quad \frac{1}{5}
\end{cases}.
\end{equation}
Under this model, it is shown that $R \geq c \log^4N$ masks, for some numerical constant $c$,  are sufficient for the convex program (\ref{LMPRf}) to uniquely recover the underlying signal up to global phase with high probability in the noiseless setting. This result has been further refined to $R \geq c \log^2N$ in \cite{gross}.

In \cite{kishoreprm}, specific deterministic masks are considered instead of random masks. In the noiseless setting, it is shown that $2$ masks are sufficient for the convex program (\ref{LMPRf}) to uniquely recover non-vanishing signals up to global phase if an $M \geq 2N$ point DFT is used. A signal $\x$ of length $N$ is said to be non-vanishing if $x[n] \neq 0$ for each $0 \leq n \leq N-1$. In particular, the 2 masks $\{\II, \D_1\}$, where $\II$ is the $N \times  N$ identity matrix and $\D_1$ is a diagonal matrix with diagonal entries given by
\begin{align}
\label{specmask}
& d_1[n] = \begin{cases}
0 \quad \quad \for n=0 \\
1 \quad \quad \for 1 \leq n \leq N-1,
\end{cases}
\end{align}
are proposed. This result has also been extended to the $N$ point DFT setup by using $5$ masks $\{\II, \D_2, \D_3, \D_4, \D_5\}$, where $\{ \D_2, \D_3, \D_4, \D_5\}$ are diagonal matrices with diagonal entries 
\begin{align}
 \nonumber
& d_2[n] = \begin{cases}
1 \quad  \for 0 \leq n \leq \lfloor{\frac{N}{2}\rfloor} \\
0 \quad \quad \otherwise
\end{cases}
\hspace{2cm} d_3[n] = \begin{cases}
1 \quad  \for 1 \leq n \leq \lfloor{\frac{N}{2}\rfloor}\\
0 \quad \quad \otherwise
\end{cases} \\
& d_4[n] = \begin{cases}
1 \quad  \for \lfloor{\frac{N}{2}\rfloor} \leq n \leq N-1 \\
0 \quad \quad \otherwise
\end{cases}
\hspace{1.25cm} d_5[n] = \begin{cases}
1 \quad  \for \lfloor{\frac{N}{2}\rfloor} +1 \leq n \leq N-1\\
0 \quad \quad \otherwise.
\end{cases} \nonumber
\end{align}
In the noisy setting, for both the aforementioned random masks and deterministic mask setup, empirical evidence strongly suggests that the recovery is stable. {\edit In the deterministic mask setup, stability guarantees are provided in \cite{kishoreprm}.}

{\em (ii) Wirtinger Flow algorithm}: An alternative recovery approach for masked signals is based on the Wirtinger flow method \cite{wirtinger}, which applies gradient descent to the least squares problem:
\begin{equation}
\min_\x \quad \sum_{r=0}^{R-1}\sum_{m=0}^{N-1} \left( Z[m,r] - \abs{\left<\f_m , \D_r \x\right>}^2 \right)^2.
\end{equation}
Minimizing such non-convex objectives is known to be NP-hard in general. Gradient descent-type methods have shown promise in solving such problems, however, their performance is very dependent on the initialization and update rules due to the fact that different initialization and update strategies lead to convergence to different (possibly local) minima.
 
Wirtinger flow (WF) is a gradient descent-type algorithm which starts with a careful initialization obtained by means of a spectral method. We refer the readers to \cite{wirtinger} for a discussion on various spectral method-based initialization strategies.
The initial estimate is then iteratively refined using particular update rules. It is argued that the average WF update is the same as the average stochastic gradient scheme update. Consequently, WF can be viewed as a stochastic gradient descent algorithm, in which only an unbiased estimate of the true gradient is observed. The authors recommend the use of smaller step-sizes in the early iterations and larger step-sizes in later iterations.
When the masks are chosen from a random model with a distribution satisfying properties similar to (\ref{distreq}), it is shown that $R \geq c \log^4 N$ masks, for some numerical constant $c$, are sufficient for the WF algorithm to uniquely recover the underlying signal up to global phase with high probability in the noiseless setting. {\edit A brief overview of WF is provided in Algorithm \ref{algo:WF}.}

\begin{algorithm}
\caption{Wirtinger Flow (WF) Algorithm\label{alg:WFA}}
\label{algo:WF}
\begin{algorithmic}
\STATE {\bf Input}: Magnitude-square measurements $\Z$ and modulations $\{ \D_0, \D_1 , \ldots , \D_{R-1}\}$, parameters $\mu_{max}$ and $t_0$
\STATE {\bf Output:} Estimate $\hat{\x}$ of the underlying signal
\STATE Initialize $\x^{(0)}$ via a spectral method (see \cite{wirtinger} for variations):  The eigenvector corresponding to the largest eigenvalue of
 \begin{equation}
\frac{1}{RN}  \left(\sum_{r=0}^{R-1} \sum_{m=0}^{N-1} Z[m,r]  \left(\D_r^\star\f_m\f_m^\star\D_r \right)  \right) \nonumber
\end{equation}
\WHILE{halting criterion false}
\STATE Update the estimate $\x^{(t+1)}$ using the rule
 \begin{equation}
\x^{(t+1)} = \x^{(t)} + \frac{\mu}{\| \x^{(0)} \| ^2} \left( \frac{1}{RN} \sum_{r=0}^{R-1} \sum_{m=0}^{N-1}  \left( Z[m,r] - \abs{\left<\f_m , \D_r \x^{(t)}\right>}^2  \right)  \D_r^\star\f_m\f_m^\star\D_r    \x^{(t)}  \right) \nonumber
\end{equation}
\STATE $\mu= \min(1 - e^{-t/t_0}, \mu_{max}) \nonumber$
\STATE $t \leftarrow t+1$
\ENDWHILE
\STATE Return $\hat{\x} \leftarrow \x^{(t)}$
\end{algorithmic}
\end{algorithm}

{\em (iii) Combinatorial methods (for the noiseless setting)}: For some choices of masks, efficient combinatorial algorithms can be employed to recover the underlying signal, where the choice of masks is closely tied to the reconstruction technique.  However, these methods are typically unstable in the presence of noise.

In \cite{kishoreprm}, a combinatorial algorithm is proposed for the $2$ masks $\{ \II, \D_1 \}$ which is shown to recover signals with $x[0] \neq 0$ up to global phase. The algorithm is derived by expressing the measurements in terms of the autocorrelation as
\begin{equation}
a_0[m] = \sum_{j=0}^{N-1-m}x[{j}]x^\star[{j+m}] \quad  \quad
a_1[m] = \sum_{j=1}^{N-1-m}x[{j}]x^\star[{j+m}] \for 0 \leq m \leq N-1. \nonumber
\end{equation}
Noting that $a_0[0] - a_1[0] = \abs{x[0]}^2$, the value of $\abs{x[0]}$ can be immediately inferred. Since $a_0[m] - a_1[m] = x[0]x^\star[m]$,  $x[m]$ for $1 \leq m \leq N-1$ is next determined up to a phase ambiguity.

Another combinatorial algorithm is proposed in \cite{candespr} for the $3$ masks $\{\mathbf{I}, \mathbf{I} + \D^s, \mathbf{I} - i\D^s\}$, where $s$ is any integer coprime with $N$ and $\D$ is a diagonal matrix with diagonal entries
\begin{equation}
d[n] = e^{{i}2\pi \frac{n}{N}} \for  0 \leq n \leq N - 1. \nonumber
\end{equation}
It is shown that signals with non-vanishing $N$ point DFT can be uniquely recovered using these masks up to global phase. Indeed, the measurements obtained in this case provide the knowledge of $\abs{y[n]}^2$, $\abs{y[n]+y[n-s]}^2$ and $\abs{y[n]-{i}y[n-s]}^2$ for $0 \leq n \leq N-1$ ($n-s$ is understood modulo $N$). Writing  $y[n] = \abs{y[n]} e^{{i}\phi[n]}$ for $0 \leq n \leq N-1$, we have 
\begin{equation}
\abs{y[n]+y[n-s]}^2 = \abs{y[n]}^2 + \abs{y[n-s]}^2 + 2 \abs{y[n]}\abs{y[n-s]} \textrm{Re}({e^{{i}{(\phi[n-s]-\phi[n])}}}), \nonumber
\end{equation}
\begin{equation}
\abs{y[n]-i y[n-s]}^2 = \abs{y[n]}^2 + \abs{y[n-s]}^2 + 2 \abs{y[n]}\abs{y[n-s]} \textrm{Im}({e^{i{(\phi[n-s]-\phi[n])}}}).\nonumber
\end{equation}
Consequently, if $y[n] \neq 0$ for $0 \leq n \leq N-1$, then the measurements provide the relative phases $\phi[n-s] - \phi[n]$ for $0 \leq n \leq N-1$. Since $s$ is coprime with $N$, by setting $\phi[0]=0$ without loss of generality, $\phi[n]$ can be inferred for $1 \leq n \leq N-1$.  Since most signals have a non-vanishing $N$ point DFT, these $3$ masks may be used to recover most signals efficiently.

{\edit In order to be able to recover all signals (as opposed to most signals), a polarization-based technique is proposed in \cite{afonso1, afonso2}. It is shown that $O(\log N)$ masks (see \cite{afonso2} for design details) are sufficient for this technique.}

In \cite{kannan2}, the authors consider a combinatorial algorithm, based on coding theoretic tools, for the $3$ masks $\{ \mathbf{I} , \mathbf{I} + \eee_0\eee_0^\star, \mathbf{I} + i \eee_0\eee_0^\star\}$, where $\eee_0$ is the $N \times 1$ column vector $(1, 0 , \ldots , 0)^T$. For signals with $x[0] \neq 0$, it is shown that the value of $\abs{x[0]}$ can be uniquely found with high probability. The phase of $x[0]$ is set to $0$ without loss of generality, and the phase of $x[n]$ relative to $x[0]$ for $0 \leq n \leq N-1$ is inferred by solving a set of algebraic equations. 
 
{\edit The results on masked phase retrieval reviewed in this section are summarized in Table II.}

\begin{figure}[t]
\begin{center}
\includegraphics[scale=0.59]{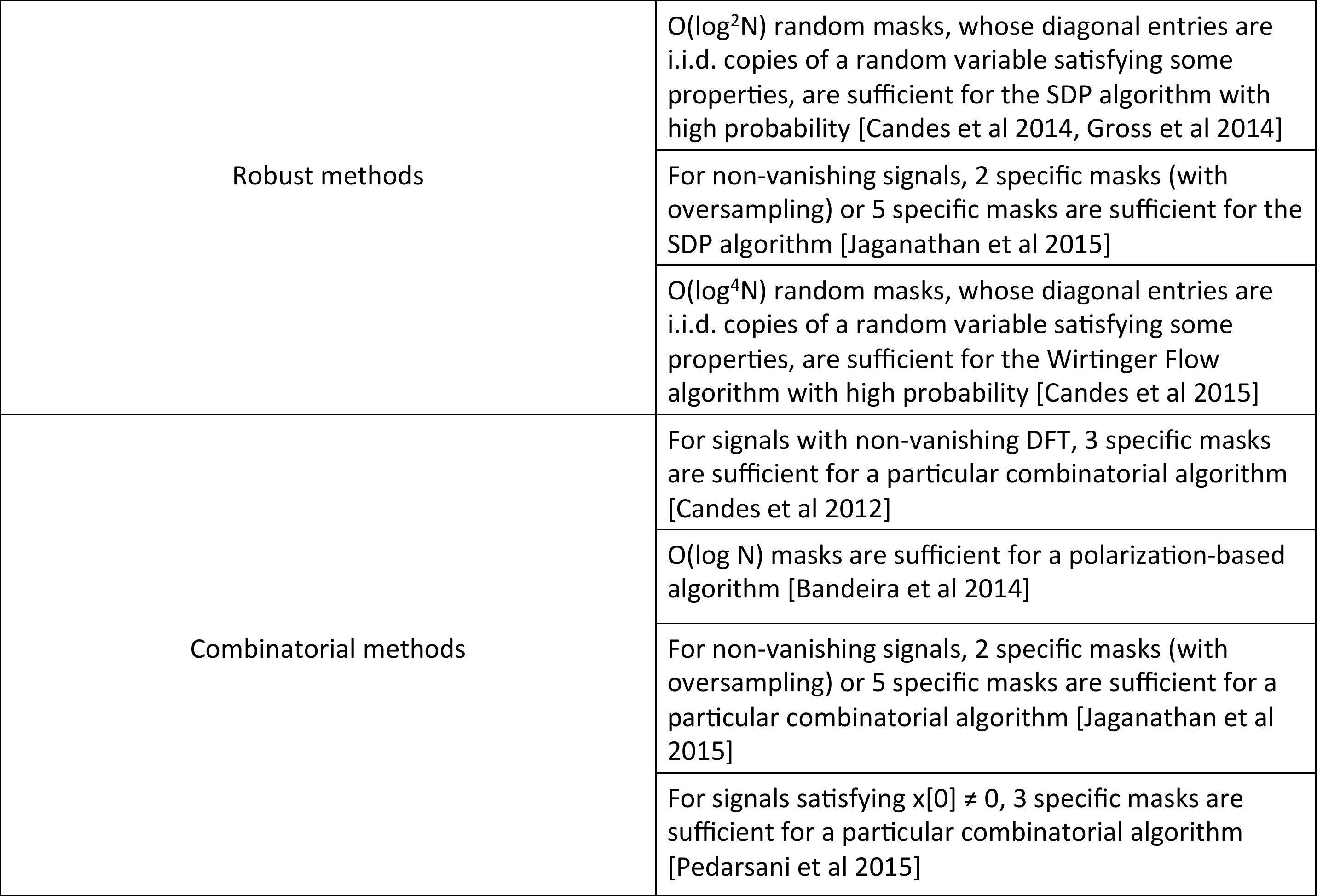}
\end{center}
\begin{center}
{Table II: Uniqueness and recovery algorithms for phase retrieval using masks.}
\end{center}
\end{figure}

%% file: PRM_Simulations.tex
\subsection{Numerical Simulations}

We now demonstrate the performance of the various algorithms discussed using numerical simulations.

\begin{figure}[t]
\begin{center}
\subfloat[{Probability of successful recovery vs $R$ using SDP.}]{\includegraphics[scale = 0.4]{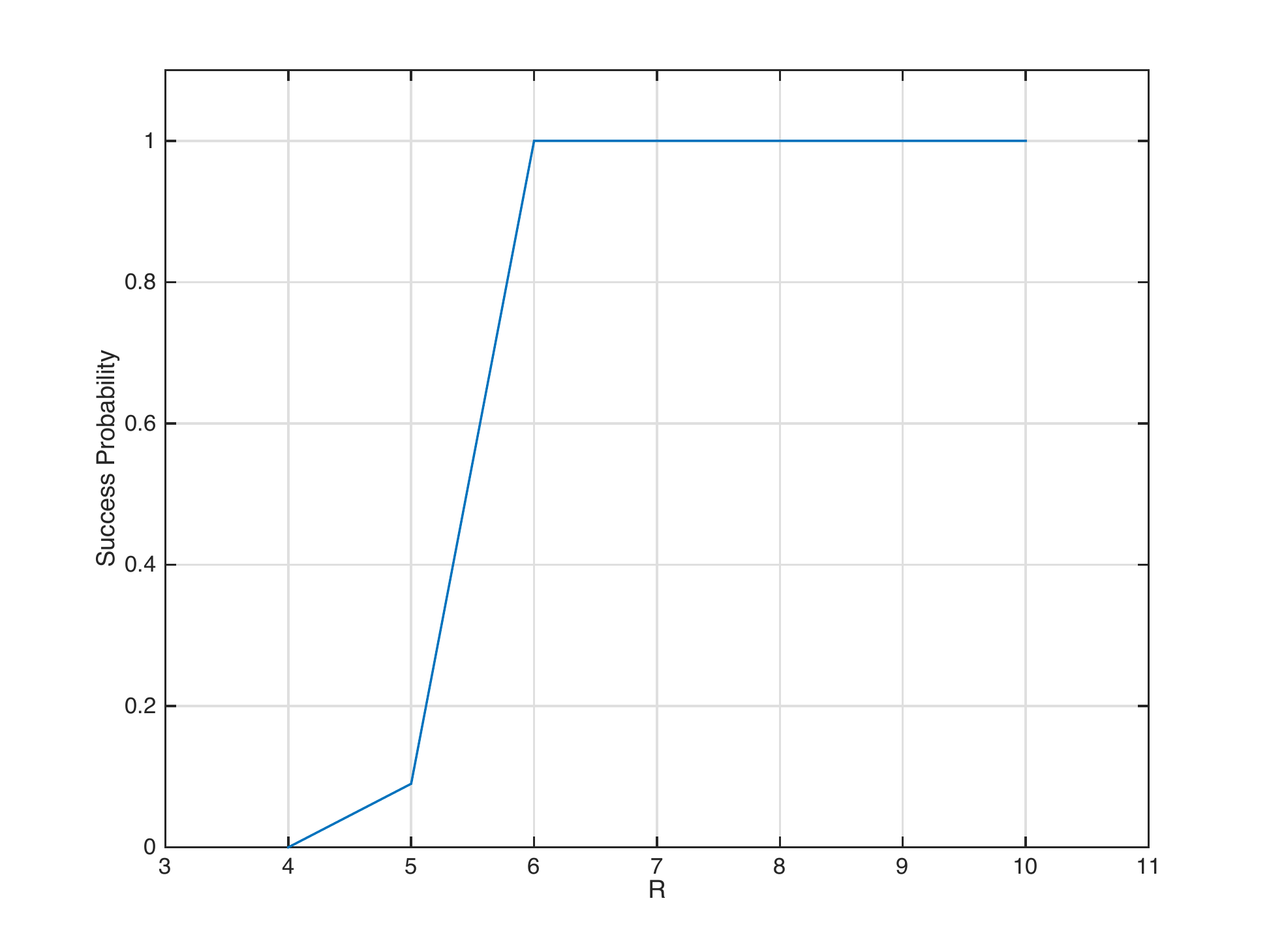}} \hspace{1cm}
\subfloat[{Probability of successful recovery vs $R$ using WF. }]{\includegraphics[scale = 0.4]{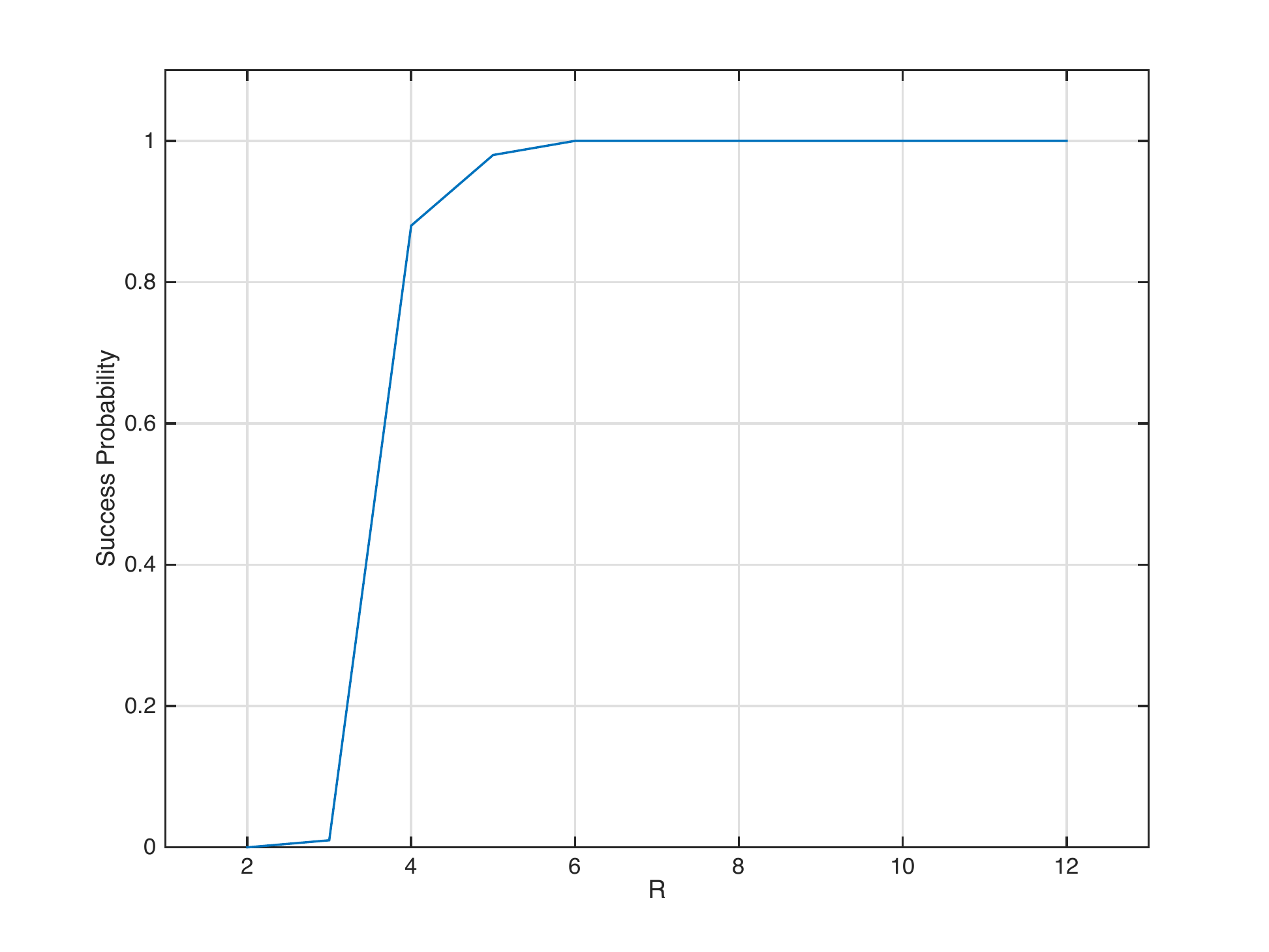}}
\end{center}
\caption{Performance of the SDP and WF algorithms in the noiseless setting for $N=128$ (courtesy of \cite{mahdi} and \cite{wirtinger}).}
\label{fig:prmsuccesssim}
\end{figure}

In the first set of simulations, we consider the performance of SDP and WF, for the random mask setup proposed in \cite{mahdi} and \cite{wirtinger} respectively.   A total of $50$ trials are performed by generating random signals of length $N=128$, such that the values in each location are chosen from an i.i.d. standard normal distribution. For the WF algorithm, the parameters $\mu_{max}$ and $t_0$ where chosen to be $0.2$ and $330$ respectively. We refer the readers to \cite{mahdi} and \cite{wirtinger} for details about the distributions used. The probability of successful recovery as a function of $R$ is plotted in Figure \ref{fig:prmsuccesssim} in the noiseless setting. It can be observed that $R \approx 6$ is sufficient for successful recovery with high probability for both methods.

\begin{figure}[t]
\begin{center}
\subfloat[{MSE (dB) vs SNR (dB) for the random mask setup proposed in \cite{mahdi} ($N=128, R = 8$).}]{\includegraphics[scale = 0.4]{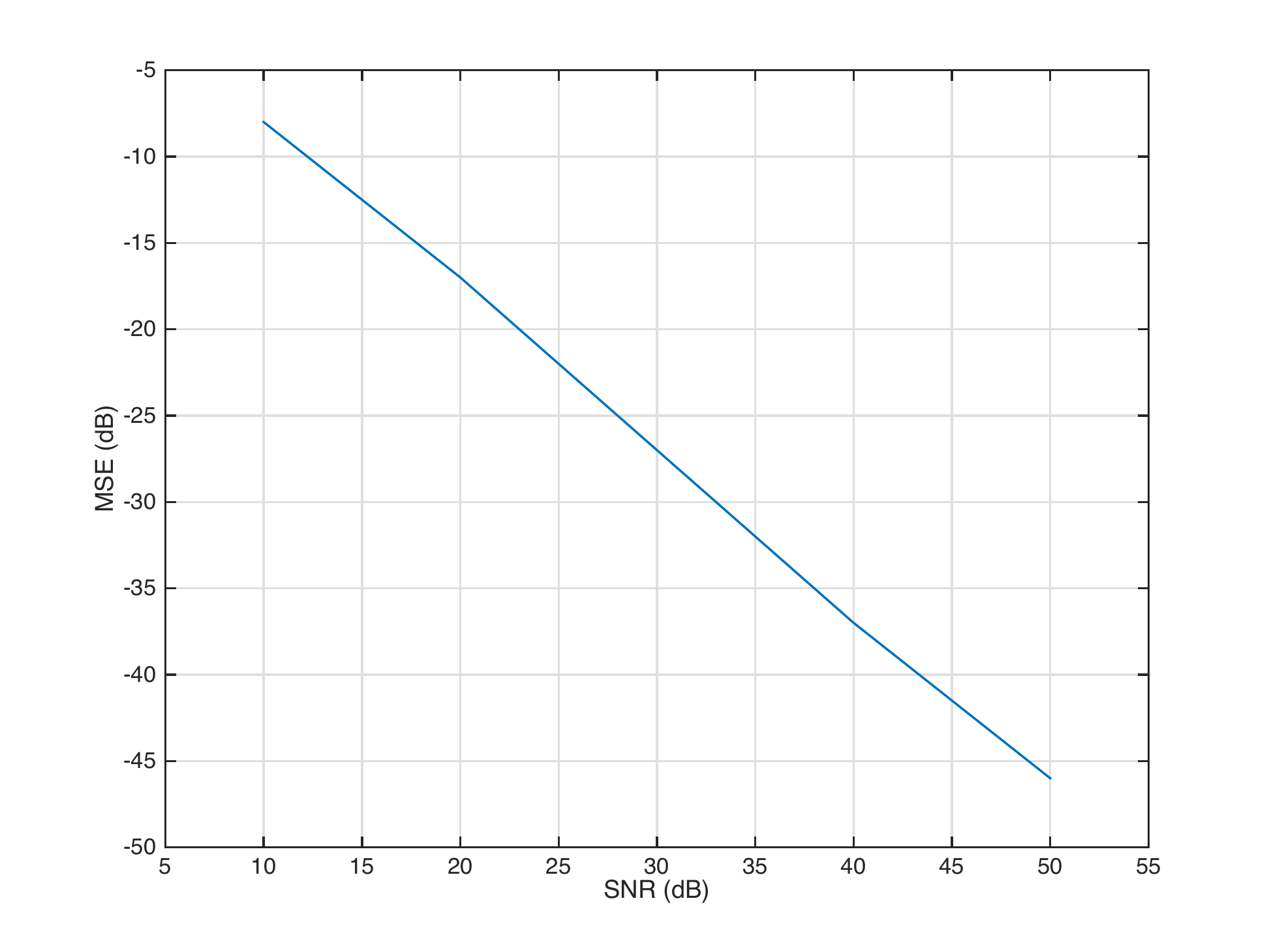}} \hspace{1cm}
\subfloat[{MSE (dB) vs SNR (dB) for the specific two-mask setup proposed in \cite{kishoreprm} ($N=32, M=64, R=2$).}]{\includegraphics[scale = 0.4]{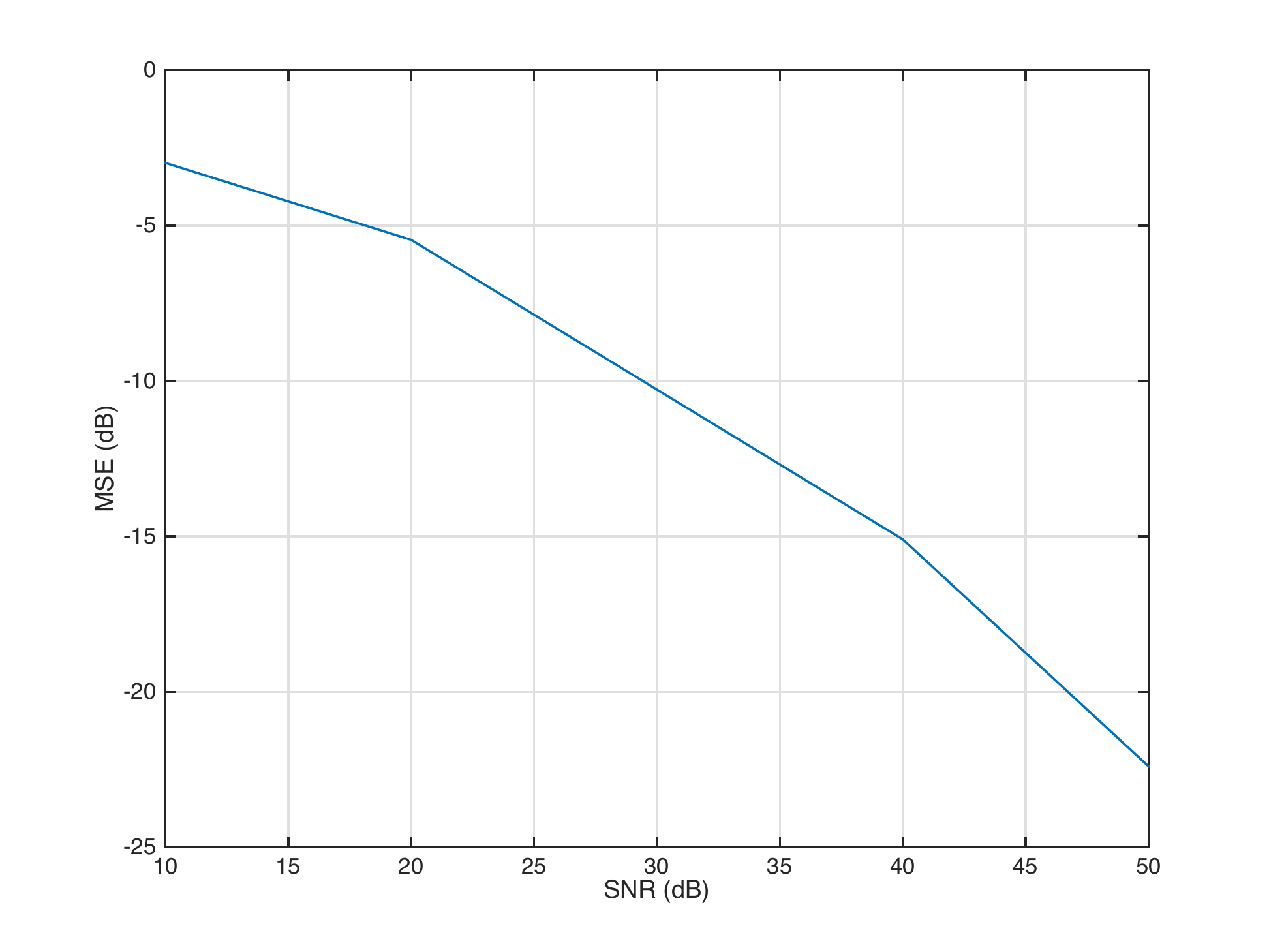}}
\end{center}
\caption{Performance of the SDP algorithm in a noisy setting (courtesy of \cite{mahdi} and \cite{kishoreprm}).}
\label{fig:prmnoisesim}
\end{figure}

In the second set of simulations, the performance of SDP, for the random mask setup proposed in \cite{mahdi} and the specific two-mask setup proposed in \cite{kishoreprm}, is evaluated in the noisy setting. In the case of random masks, eight masks are used and the rest of the parameters are the same as in Figure \ref{fig:prmsuccesssim}. For the two-mask setup, $N=32$ is chosen and the $64$ point DFT is considered. The normalized mean-squared error is plotted as a function of SNR for these two settings in Figure \ref{fig:prmnoisesim}. While a direct comparison of the two methods is meaningless since the first one uses four times as many measurements, the results clearly show that the recovery is stable in the presence of noise.

%\begin{equation}
%MSE = \min\limits_{\abs{c} =1 } \frac{ \|\hat{\x} - c \x_0\|^2 } { \| \x_0 \|^2 },
%\label{nmse}
%\end{equation}

%% file: STFTPR_Introduction.tex
\section{STFT Phase Retrieval}

In this section we consider introducing redundancy into the Fourier measurements by using the short-time Fourier transform (STFT).
The key idea is to introduce redundancy in the magnitude-only measurements by maintaining a substantial overlap between adjacent short-time sections.
As we will see, the redundancy offered by the STFT enables unique and robust recovery in many cases. Furthermore, using the STFT leads to improved performance over recovery from the oversampled Fourier magnitude with the same number of measurements.

{\edit Phase retrieval from the STFT magnitude
has been used in several signal processing applications, for example in speech and audio processing \cite{nawab,lim},
where the spectral content of speech changes over time \cite{nawab, oppenheim}.
It has also been applied extensively in optics. One example is in
frequency resolved optical gating (FROG) or XFROG which are used for
characterizing ultra-short laser pulses by optically producing the STFT magnitude of the measured pulse \cite{trebino, kane}.
In FROG the pulse itself (or a function of the pulse) is used to gate the measured signal while in XFROG gating is performed by a fixed known window.
Another example is ptychographical CDI \cite{fienupnew} or Fourier ptychography \cite{humphry, rodenburg, yang}, a technology which has enabled X-ray, optical and electron microscopy with increased spatial resolution without the need for advanced lenses.}

Let $\w = ( w[0] , w[1] , \ldots , w[W-1] )$ be a window of length $W$ such that it has non-zero values only within the interval $[0,W-1]$. The STFT of  $\x$ with respect to $\w$, denoted by $\Y_w$, is defined as
\begin{equation}
Y_w[ m , r ] = \sum_{n = 0}^{N-1} x[n]w[rL - n ] e^{- i 2 \pi \frac{ m n } { N } } \for 0 \leq m \leq N - 1\aand 0 \leq r \leq R-1,
\end{equation}
where the parameter $L$ denotes the separation in time between adjacent short-time sections and the parameter $R=\lceil\frac{N+W-1}{L}\rceil$ denotes the number of short-time sections considered.

\begin{figure}
\begin{center}
\includegraphics[scale=0.45]{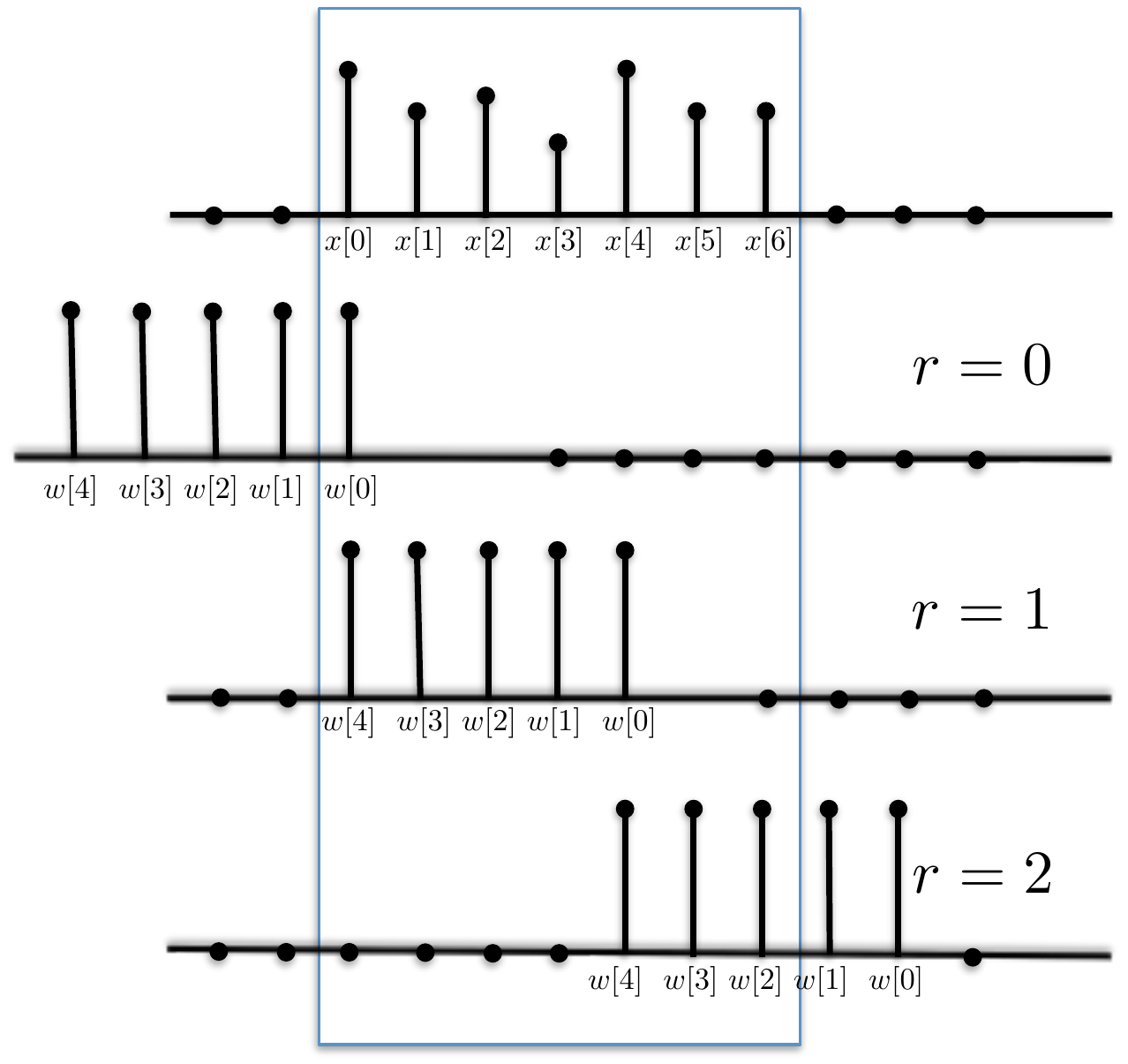} \includegraphics[scale=0.45]{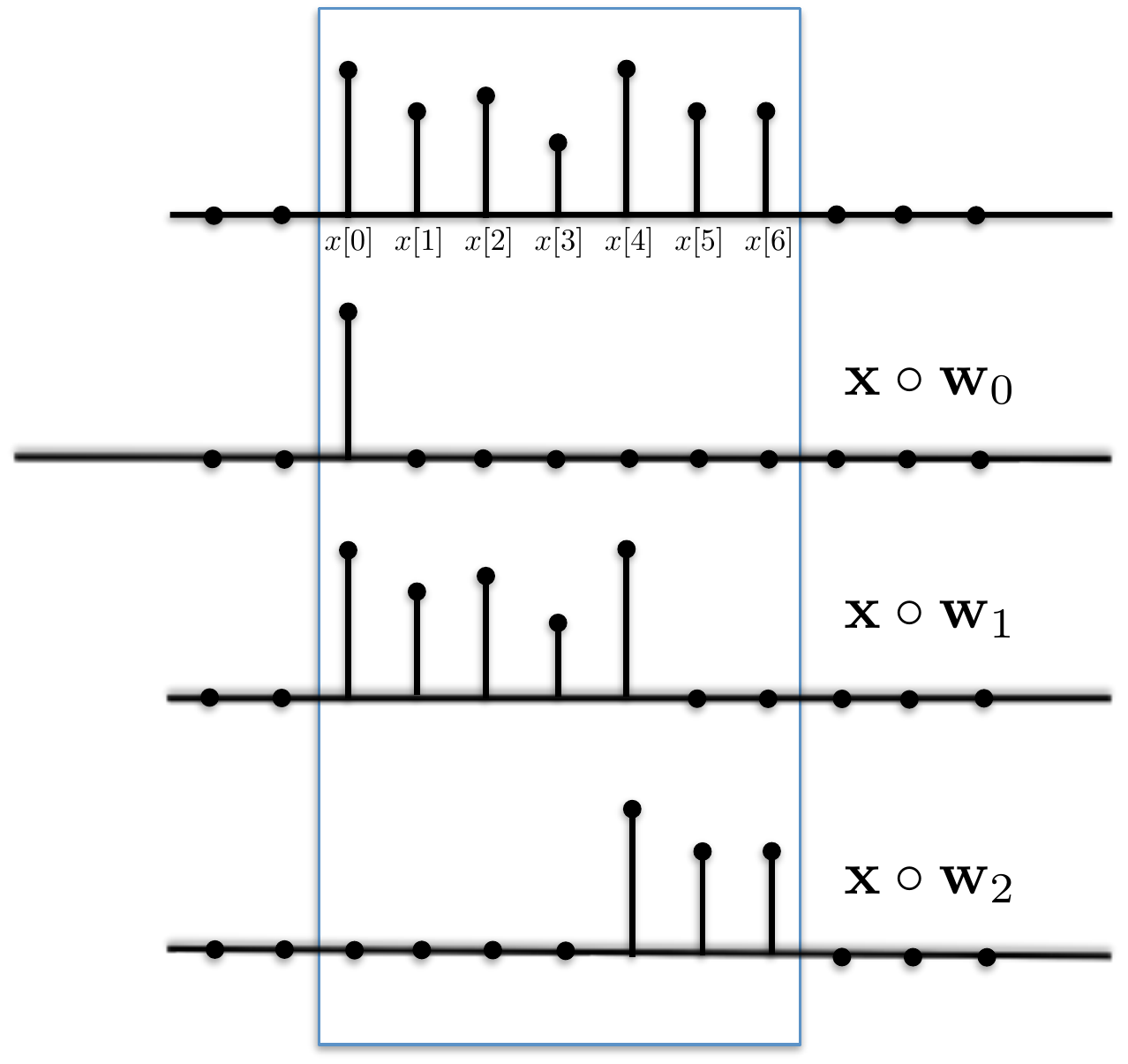}
\end{center}
\caption{Sliding window interpretation of the STFT for $N=7$, $W = 5$ and $L=4$. The shifted window overlaps with the signal for $3$ shifts, and hence $R=3$ short-time sections are considered {\edit (courtesy of \cite{kishorestft})}.}
\label{fig:STFTexample}
\end{figure}

The STFT can be interpreted as follows: Suppose $\w_r$ denotes the signal obtained by shifting the flipped window $\w$ by $rL$ time units (i.e., $w_r[n] = w[rL - n]$) and let $\circ$ denote the Hadamard (element-wise) product operator. The $r${th} column of  $\Y_w$, for $0 \leq r \leq R - 1$, corresponds to the $N$ point DFT of $\x \circ \w_r$. In essence, the window is flipped and slid across the signal (see Figure \ref{fig:STFTexample} for a pictorial representation), and $\Y_w$ corresponds to the Fourier transform of the windowed signal recorded at regular intervals. This interpretation is known as the {\em sliding window} interpretation \cite{schafer}.

The problem of reconstructing a signal from its STFT magnitude is known as STFT phase retrieval. In fact, STFT phase retrieval can be viewed as a special instance of phase retrieval using masks, where the different masks considered are time-shifted copies of the window. This can be seen as follows: Let $\Z_w$ be the $N \times R$ measurements corresponding to the magnitude-square of the STFT of $\x$ with respect to $\w$ so that $Z_w[m,r] = \abs{Y_w[m,r]}^2$. Let $\W_r$, for $0 \leq r \leq R-1$, be the $N \times N$ diagonal matrix with diagonal elements $( w_r[0], w_r[1], \ldots , w_r[N-1]) $. STFT phase retrieval can then be mathematically stated as
\begin{align}
\label{STFTPRf}
&\textrm{find} \hspace{2cm} \x  \\
\nonumber & \textrm{subject to} \hspace{1cm} Z_w[m,r] = \abs{\left<\f_m,\W_r\x\right>}^2 \for 0 \leq m \leq N - 1\aand 0 \leq r \leq R-1,
\end{align}
which is equivalent to \eqref{MPRf}.

%% file: STFTPR_Uniqueness.tex
\subsection{Uniqueness}

In this subsection, we review the main results regarding uniqueness of STFT phase retrieval. As before, a signal $\x$ is non-vanishing if $x[n] \neq 0$ for each $0 \leq n \leq N-1$. Similarly, a window $\w$ is called non-vanishing if $w[n] \neq  0$ for all $0 \leq n \leq W-1$. These results are summarized in Table III.

First, we argue that, for $W<N$ (which is typically the case), $L<W$ is a necessary condition in order to be able to uniquely identify most signals: If $L > W$, then the STFT magnitude does not contain any information from some locations of the signal, because of which most signals cannot be uniquely identified. If $L=W$, then the adjacent short-time sections do not overlap and hence STFT phase retrieval is equivalent to a series of non-overlapping phase retrieval problems. Consequently, as in the case of phase retrieval, most 1D signals are not uniquely identifiable. For higher dimensions (2D and above), almost all windowed signals corresponding to each of the short-time sections are uniquely identified up to trivial ambiguities if  $\w$ is non-vanishing. However, since there is no way of establishing relative phase, time-shift or conjugate-flip between the windowed signals corresponding to the various short-time sections, most signals cannot be uniquely identified. For example, suppose we choose $L=W=2$ and $w[n] = 1$ for all $0 \leq n \leq W-1$. Consider the signal $\x_1 = ( 1 , 2 , 3 )^T$ of length $N=3$. Signals $\x_1$ and $\x_2 = ( 1 , -2 , -3 )^T$ have the same STFT magnitude. In fact, more generally, signals $\x_1$ and $( 1 , e^{{i} \phi}2 , e^{{i} \phi}3 )^T$, for any $\phi$, have the same STFT magnitude.

\subsubsection{Non-vanishing signals}

In \cite{kishorestft} it is shown that most non-vanishing signals are uniquely identifiable from their STFT magnitude up to global phase if $1 \leq L<W$ and $\w$ is chosen such that it is non-vanishing with $W \leq \frac{N}{2}$. In other words, this result states that, if adjacent short-time sections overlap (the extent of overlap does not matter), then STFT phase retrieval is almost surely well posed under mild conditions on the window. Observe that, like in phase retrieval, the global phase ambiguity cannot be resolved due to the fact that signals $\x$ and $e^{{i} \phi} \x$, for any $\phi$, have the same STFT magnitude regardless of the choice of $\{\w, L\}$. However, unlike in phase retrieval, the time-shift and conjugate-flip ambiguities are resolvable for most non-vanishing signals.

For some specific choices of $\{\w,L\}$, all non-vanishing signals are uniquely identified from their STFT magnitude up to global phase. In \cite{eldarstft}, it is shown that the STFT magnitude can uniquely identify non-vanishing signals up to global phase for $L=1$ if the window $\w$ is chosen such that the $N$ point DFT of $(|w[0]|^2, w[1]^2, \ldots, |w[N-1]|^2)$ is non-vanishing, $2 \leq W  \leq \frac{N+1}{2}$ and $W-1$ is coprime with $N$. An alternative result is that if the first $L$ samples are known a priori, then the STFT magnitude can uniquely identify non-vanishing signals for any $L$ as long as the window $\w$ is chosen such that it is non-vanishing and $2L \leq W \leq \frac{N}{2}$ \cite{nawab}.

\subsubsection{Sparse signals}

While the aforementioned results provide guarantees for non-vanishing signals, they do not say anything about sparse signals. In this section, a signal $\x$ of length $N$ is said to be sparse if $x[n] = 0$ for at least one $0 \leq n \leq N-1$. Based on intuitions from compressed sensing, one might expect to recover sparse signals easier than non-vanishing signals. However, this is actually not the case for STFT phase retrieval.

The following example is provided in \cite{eldarstft} to demonstrate that the time-shift ambiguity cannot be resolved for some classes of sparse signals and some choices of $\{\w, L\}$: Suppose $L \geq 2$, $W$ is a multiple of $L$ and $w[n] = 1$ for all $0 \leq n \leq W-1$. Consider a signal $\x_1$ of length $N \geq L+1$ such that it has non-zero values only within an interval of the form $[(t -1)L + 1, (t-1)L+L-p]  \subset [0,N-1]$ for some integers $1 \leq p \leq L-1$ and $t \geq 1$. The signal $\x_2$ obtained by time-shifting $\x_1$ by $q \leq p$ units (i.e., $x_2[i] = x_1[i - q]$) has the same STFT magnitude. The issue with this class of sparse signals is that the STFT magnitude is identical to the Fourier magnitude because of which time-shift and conjugate-flip ambiguities cannot be resolved.

It is further shown that there are sparse signals that are not recoverable even up to trivial ambiguities for some choices of  $\{\w, L\}$. Consider two non-overlapping intervals $[u_1, v_1] , [u_2,v_2] \subset [0, N-1]$ such that $u_2 - v_1 > W$, and choose signals $\x_1$ supported on $[u_1,v_1]$ and $\x_2$ supported on $[u_2,v_2]$. The magnitude-square of the STFT of $\x_1+\x_2$ and of $\x_1-\x_2$ are equal for any choice of $L$. For such examples of sparse signals the two intervals with non-zero values are separated by a distance greater than $W$ because of which there is no way of establishing relative phase using a window of length $W$.

The examples above establish the fact that sparse signals are harder to recover than non-vanishing signals from their STFT magnitude. Since the issues are primarily due to a large number of consecutive zeros, the uniqueness guarantees for non-vanishing signals have been extended to incorporate sparse signals with limits on the number of consecutive zeros. In \cite{kishorestft}, it is shown that most sparse signals with less than $\min\{W-L, L\}$ consecutive zeros are uniquely identified from their STFT magnitude up to global phase and time-shift ambiguities if adjacent short-time sections overlap (i.e., $L < W$) and $\w$ is chosen such that it is non-vanishing and $W \leq \frac{N}{2}$. The work in \cite{nawab} establishes that if $L$ consecutive samples are known a priori, starting from the first non-zero sample, then the STFT magnitude can uniquely identify signals with at most $W-2L$ consecutive zeros for any $L$ if the window $\w$ is chosen such that it is non-vanishing and $2L \leq W \leq \frac{N}{2}$. 

\begin{figure}[t]
\begin{center}
\includegraphics[scale=0.59]{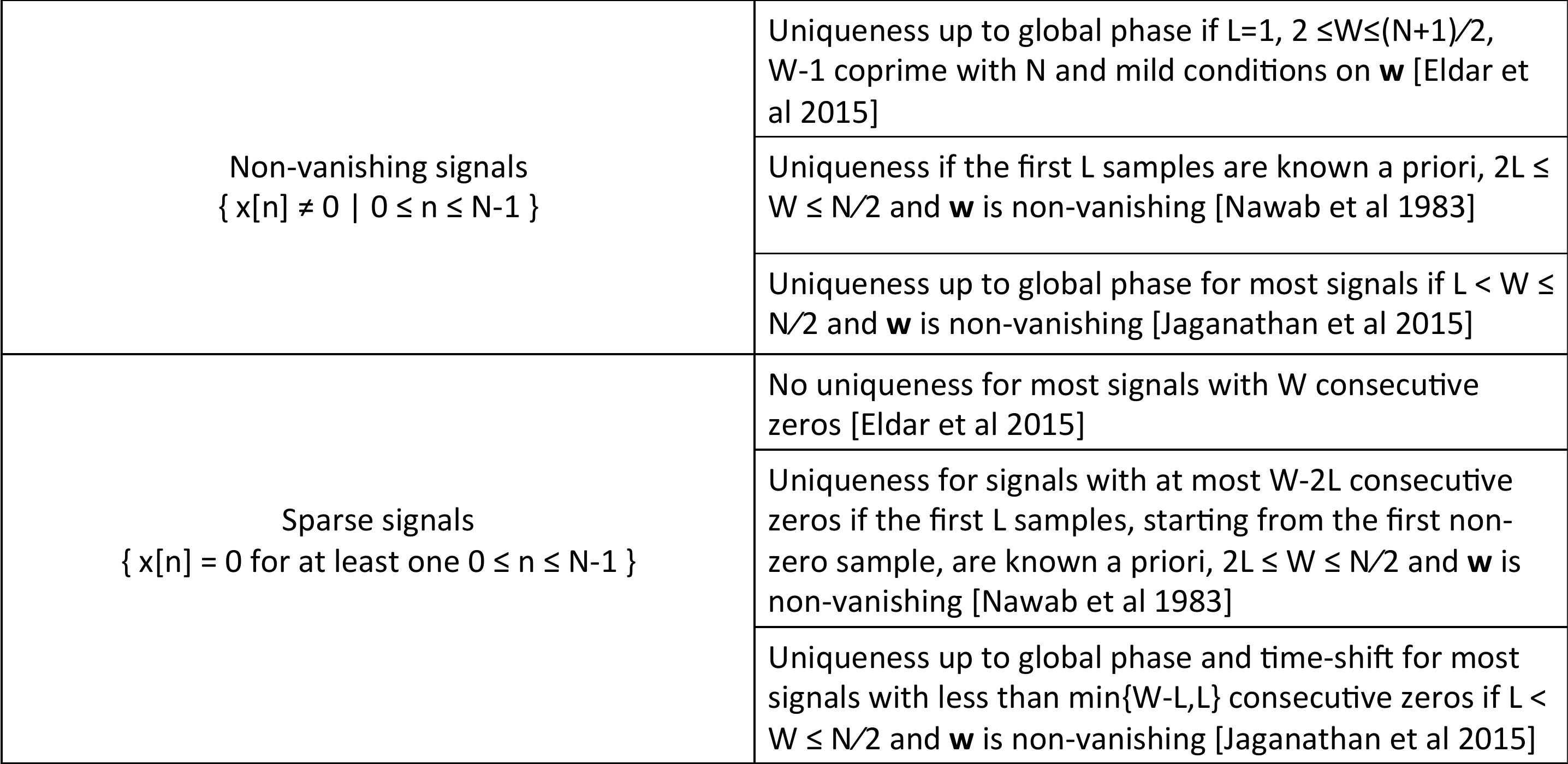}
\end{center}
\begin{center}
{Table III: Uniqueness of STFT phase retrieval.}
\end{center}
\end{figure}

%% file: STFTPR_Algorithms.tex
\subsection{Algorithms}

Phase retrieval techniques based on alternating projections, SDP and greedy methods (e.g., GESPAR for sparse signals) have been modified to solve STFT phase retrieval efficiently and robustly.  In this subsection, we provide a survey of the existing STFT phase retrieval algorithms.

{\em (i) Alternating projections}: The classic alternating projection method has been adapted to solve STFT phase retrieval by Griffin and Lim \cite{lim}. To this end, STFT phase retrieval is formulated as the following least-squares problem:
\begin{equation}
\label{STFTleastsquares}
\min_\x \quad \sum_{r=0}^{R-1}  \sum_{m=0}^{N-1} { \left( Z_w[ m , r ] - \abs{\left<\f_m,\W_r\x\right>}^2 \right)^2 }.
\end{equation}
The Griffin-Lim (GL) algorithm attempts to minimize this objective by starting with a random initialization and imposing the time domain and STFT magnitude constraints alternately using projections. The details of the various steps are provided in Algorithm \ref{alg:GLA}. The objective is shown to be monotonically decreasing as the iterations progress. In the noiseless setting, an important feature of the GL method is its empirical ability to converge to the global minimum when there is substantial overlap between adjacent short-time sections. However, no theoretical recovery guarantees are available. In the noisy setting, the algorithm has the same limitations as the GS and HIO techniques for standard phase retrieval. 

In optics, a slight variation of the GL iterations is used, referred to as principal components generalized projections (PCGP). We refer the readers to \cite{kane} for details.

%Compute $\x^{(\ell)}$ such that the windowed version of $\x^{(\ell)}$ (i.e., $\x^{(\ell)} \circ \w_r$ for each $0 \leq r \leq R-1$) is closest to $\x'^{(\ell)}_r$ for each $0 \leq r \leq R-1$ (in a least square sense)

\begin{algorithm}
\caption{Griffin-Lim (GL) Algorithm\label{alg:GLA}}
\begin{algorithmic}
\STATE {\bf Input}: STFT magnitude-square measurements $\Z_w$ and window $\w$
\STATE {\bf Output:} Estimate $\hat{\x}$ of the underlying signal
\STATE {\bf Initialize}: Choose a random input signal $\x^{(0)}$, $\ell = 0$
\WHILE{halting criterion false}
\STATE $\ell \leftarrow \ell+1$
\STATE Compute the STFT of $\x^{(\ell-1)}$: $Y_w^{(\ell)}[m,r]=\sum_{n=0}^{N-1} x^{(\ell-1)}[n]w[rL-n]e^{-{i}2\pi \frac{m n}{N}}$
\STATE Impose STFT magnitude constraints: $Y'^{(\ell)}_w[m,r]=\frac{Y_w^{(\ell)}[m,r]}{ \abs{Y_w^{(\ell)}[m,r]} }\sqrt{Z_w[m,r]}$
\STATE Compute the inverse DFT of $\Y'^{(\ell)}_w$ for each short-time section to obtain windowed signals $\x'^{(\ell)}_r$
\STATE  Impose time domain constraints to obtain $\x^{(\ell)}$: $x^{(\ell)}[n] = \frac{\sum_r  x'^{(\ell)}_r [n]w^\star[rL-n]  }{\sum_r \abs{w[rL-n]}^2 }$
\ENDWHILE
\STATE return $\hat{\x} \leftarrow {\x}^{(\ell)}$
\end{algorithmic}
\end{algorithm}

%However, since the objective is non-convex,  different initializations can lead to different solutions and there are no theoretical guarantees available on convergence to a global optimum. As the performance is dependent on initial points, it is possible and recommended to try several initializations.

{\em (ii) SDP method}: In \cite{sun, kishorestft}, the SDP-based phase retrieval approach has been applied to STFT phase retrieval, leading to the STliFT algorithm detailed in Algorithm \ref{STFTLIFTalgo}.

\begin{algorithm}[h]
\caption{STliFT}\label{STFTLIFTalgo}
\begin{algorithmic}
\STATE {\bf Input}: STFT magnitude-square measurements $\Z_w$ and window $\w$\\
\STATE {\bf Output}: Estimate $\hat{\x}$ of the underlying signal
\STATE Obtain $\hat{\X}$ by solving
\STATE \begin{align}
\label{STFTPRR}
&\textrm{minimize} \hspace{1cm} \mbox{trace}( \X )  \\
\nonumber & \textrm{subject to} \hspace{0.8cm} {Z_w[m,r]} =  \mbox{trace}(~ \f_m \f_m^\star ~(\X \circ {\w}_r{\w}_r^\star) ~)
\for  0 \leq m \leq N-1 \aand 0 \leq r \leq R -1,\\
& \nonumber \hspace{2.4cm}  \X \succcurlyeq 0.
\end{align}
\STATE Return $\hat{\x}$, where $\hat{\x} \hat{\x}^\star$ is the best rank-one approximation of $\hat{\X}$
\end{algorithmic}
\end{algorithm}

Many recent results in related problems like {\em generalized phase retrieval} \cite{candespl} and {\em phase retrieval using random masks} \cite{mahdi, gross} suggest that one can provide conditions on $\{\w, L\}$, which when satisfied, ensure that the SDP-based algorithm  correctly recovers the underlying signal. In \cite{kishorestft}, it is shown that STliFT uniquely recovers non-vanishing signals from their STFT magnitude up to global phase if $L=1$, $2 \leq W \leq \frac{N}{2} $ and $\w$ is non-vanishing. It is further proved that STliFT uniquely recovers non-vanishing signals from their STFT magnitude for any $2 \leq 2L \leq W \leq \frac{N}{2} $ and non-vanishing $\w$ if the first $\frac{L}{2}$ samples are known a priori.
These guarantees only partially explain the excellent empirical performance of STliFT. In the noiseless setting, STliFT is observed empirically to recover most signals for any $2 \leq 2L \leq W \leq \frac{N}{2}$. In the presence of noise, like many SDP methods, STliFT appears to exhibit stable recovery. This behaviour is demonstrated in the numerical simulation section.

{\em (iii) GESPAR for sparse signals}: If it is known a priori that the underlying signal is sparse, then the recovery performance can potentially be improved by exploiting this sparsity. One method for taking sparsity into account is by adapting GESPAR \cite{eldargespar} to STFT phase retrieval \cite{eldarstft}. Simulations demonstrate that GESPAR is able to exploit both the redundancy in the measurements and the sparsity of the input, leading to high probability of successful reconstruction and stable recovery in the presence of noise as long as sufficient redundancy is introduced into the measurement process. We demonstrate some of these results via simulations below. 

{\em (iv) Combinatorial methods (for the noiseless setting):} For some choices of $\{\w, L \}$, researchers have developed efficient combinatorial methods specific to the STFT phase retrieval setup. In \cite{nawab}, a sequential reconstruction technique is proposed for non-vanishing signals when $L=1, W \geq 2$. The algorithm works as follows: $x[0]$ is reconstructed up to a phase from the measurement corresponding to the short-time section $r=0$. Using the knowledge of $x[0]$, $x[1]$ is reconstructed up to a relative phase from the measurement corresponding to the short-time section $r=1$. Continuing sequentially, the entire signal can be estimated up to global phase (see \cite{nawab} for details). In \cite{eldarstft}, a reconstruction algorithm is proposed where $\abs{x[n]}$ for $0 \leq n \leq N-1$ is first reconstructed by solving a linear system of equations and the phases of $x[n]$ are then sequentially reconstructed using a combinatorial method.

Like many sequential reconstruction approaches, these methods are typically unstable in the noisy setting due to error propagation.

%\begin{thm}
%STliFT uniquely recovers (up to global phase) a nowhere-vanishing signal $\x$ from its STFT magnitude measurements if
%\begin{enumerate}
%\item $L = 1$, $2 \leq W \leq \frac{n}{2} $
%\item $w[0]w[1] \neq 0$.
%\end{enumerate}
%\label{l1thm}
%\end{thm}

%Under the conditions of Theorem \ref{l1thm}, one could devise a simple combinatorial algorithm based on sequential reconstruction to recover the true signal. The measurement corresponding to $m=0$ is  $a_w[0,0] = |w[0]x[0]|^2$, from which $x[0]$ can be recovered up to a phase. The measurements corresponding to $m=1$ are $a_w[1,0] = |w[1]x[0]|^2 + |w[0]x[1]|^2$ and $a_w[1,1] = w[1]x[0]w^\star[0]x^\star[1]$. Using the knowledge of $x[0]$ up to a phase, $x[1]$ can be estimated up to the same phase. Continuing sequentially, the entire signal $\x$ can be estimated up to a global phase (see \cite{nawab} for details).

%\begin{thm}
%STliFT uniquely recovers (up to global phase) a nowhere-vanishing signal $\x$ from its STFT magnitude measurements if
%\begin{enumerate}
%\item $2 \leq 2L \leq W \leq \frac{n}{2} $
%\item $x[i]$ for $0 \leq i \leq \lfloor{\frac{L}{2}\rfloor}$ known a priori
%\item $\w$ is nowhere-vanishing.
%\end{enumerate}
%\label{lhalfthm}
%\end{thm}

%Similarly, under the conditions of Theorem \ref{lhalfthm}, one could devise a simple combinatorial algorithm based on sequential reconstruction to recover the true signal (see \cite{nawab} for details). However, the STliFT is extremely robust compared to sequential reconstruction methods in the presence of measurement noise.

%% file: STFTPR_Simulations.tex
\subsection{Numerical Simulations}

We now demonstrate the performance of various STFT phase retrieval algorithms using numerical simulations.

\begin{figure}
\begin{center}
\subfloat[{Probability of successful recovery in the noiseless setting (white region: success with probability $1$, black region: success with probability $0$).}]{\includegraphics[scale = 0.5]{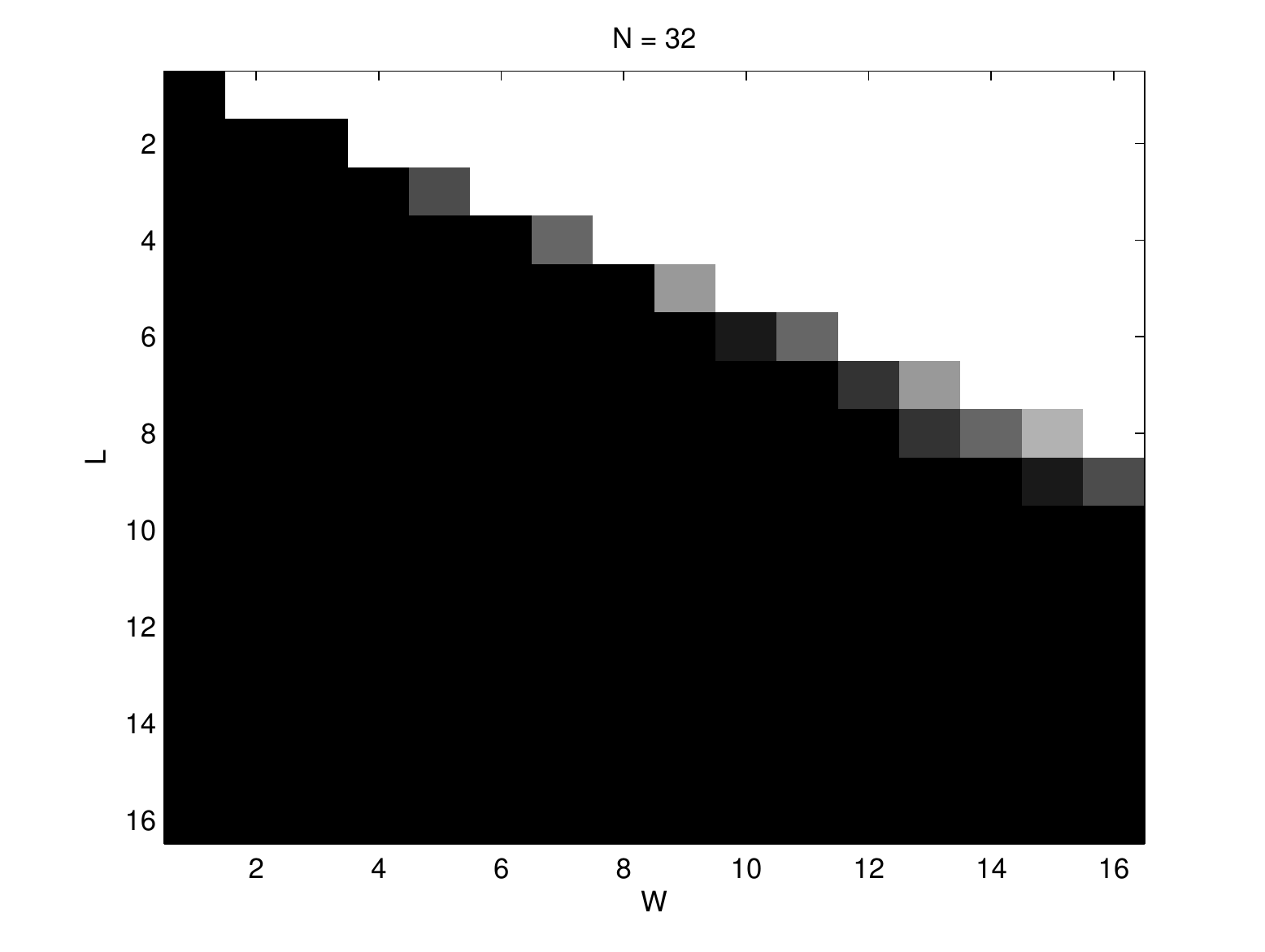} \label{fig:successprobabilitystft}
}  \hspace{1cm}
\subfloat[ {MSE (dB) vs SNR (dB) in the noisy setting.}]{\includegraphics[scale = 0.4]{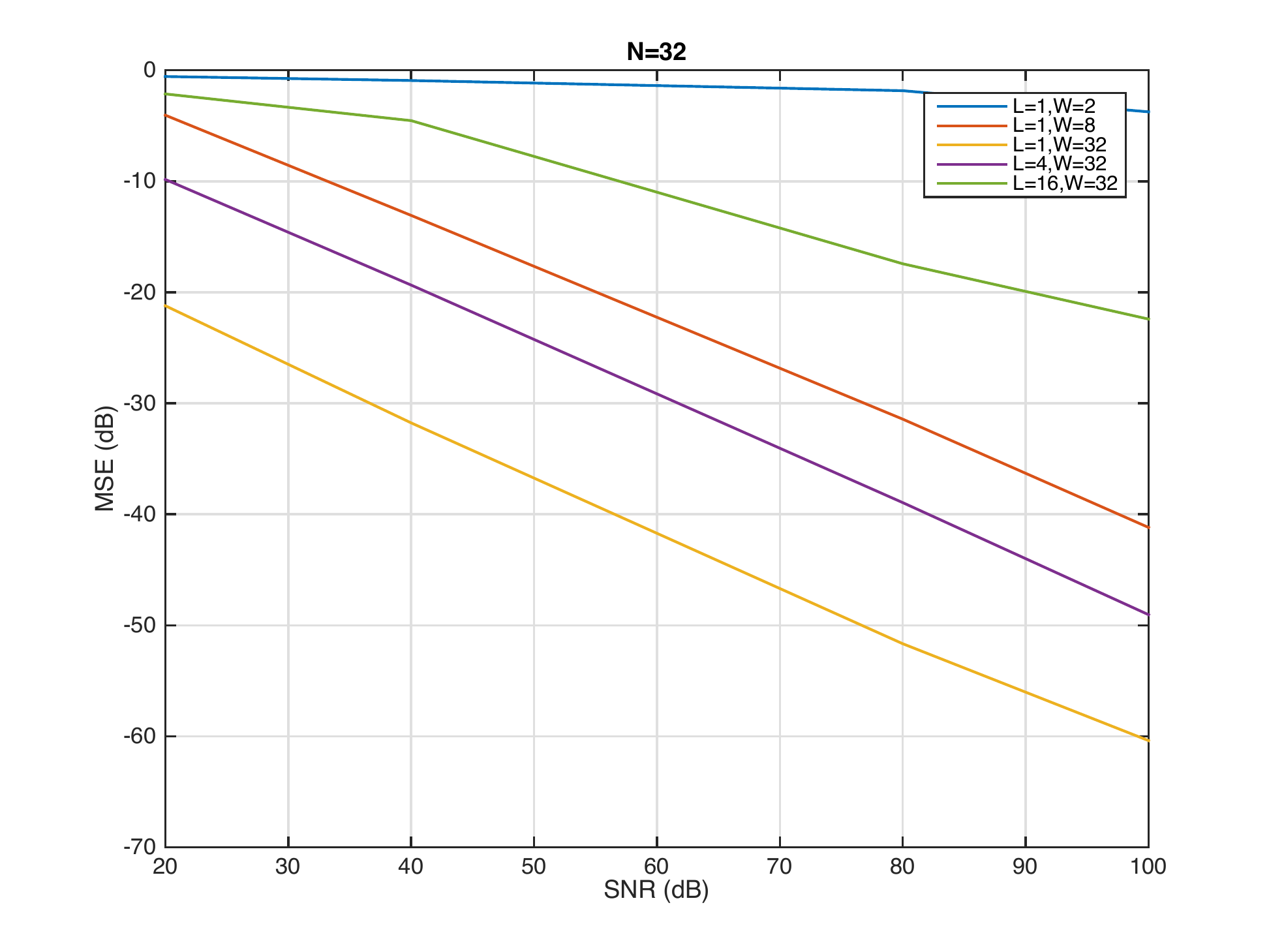} \label{fig:errorstft}}
\end{center}
\caption{Performance of STliFT for $N=32$ and various choices of $\{W,L\}$ (courtesy of \cite{kishorestft}).}
\end{figure}

In the first set of simulations, the non-vanishing signal recovery ability of STliFT is evaluated for $N=32$ in the noiseless and noisy settings. The window $\w$ is chosen such that $w[n]=1$ for all $0 \leq n \leq W-1$ and the parameters $L$ and $W$ are varied between $1$ and $\frac{N}{2}$. For each choice of $\{W,L\}$, $100$ simulations are performed by randomly choosing non-vanishing signals, such that the values in each location are drawn from an i.i.d. standard normal distribution. The probability of successful recovery as a function of $\{W,L\}$ in the noiseless setting is shown in Figure \ref{fig:successprobabilitystft}. Observe that STliFT successfully recovers the underlying signal with very high probability if $2L \leq W \leq \frac{N}{2}$. The case $L = \frac{N}{4},W = \frac{N}{2}$ uses only {\em six} windowed measurements and the underlying signal is recovered with very high probability by STliFT. Given the limited success of SDP-based methods in the Fourier phase retrieval setup, this is very encouraging.
In the noisy setting, the normalized mean square error is plotted in Figure \ref{fig:errorstft} for various choices of SNR and $\{W,L\}$. The linear relationship between $\log( \textrm{MSE})$ and SNR indicates that STliFT stably recovers the underlying signal. It can also be seen that when there is significant overlap between adjacent short-time sections, the signals are recovered more stably, which is not surprising.

In the second set of simulations, the sparse signal recovery abilities of STFT-GESPAR, GL algorithm and PCGP are evaluated for $N=64$ in the noiseless and noisy settings. The window satisfies $w[n]=1$ for $0 \leq n \leq W-1$ and the parameters $L$ and $W$ are chosen to be $\{2,4,8,16\}$ and $16$ respectively. The sparsity dictionary is a random basis with i.i.d. standard normal variables, followed by normalization of the columns. To generate the sparse inputs, for each sparsity level $k$, the $k$ locations for the non-zero values are chosen uniformly at random. The signal values over the selected support are then drawn from an i.i.d. standard normal distribution.

\begin{figure}
\begin{center}
\subfloat[{Probability of successful recovery vs sparsity in the noiseless setting for varying measurements (i) 64, (ii) 128, (iii) 256, (iv) 512.}]{\includegraphics[scale=0.31]{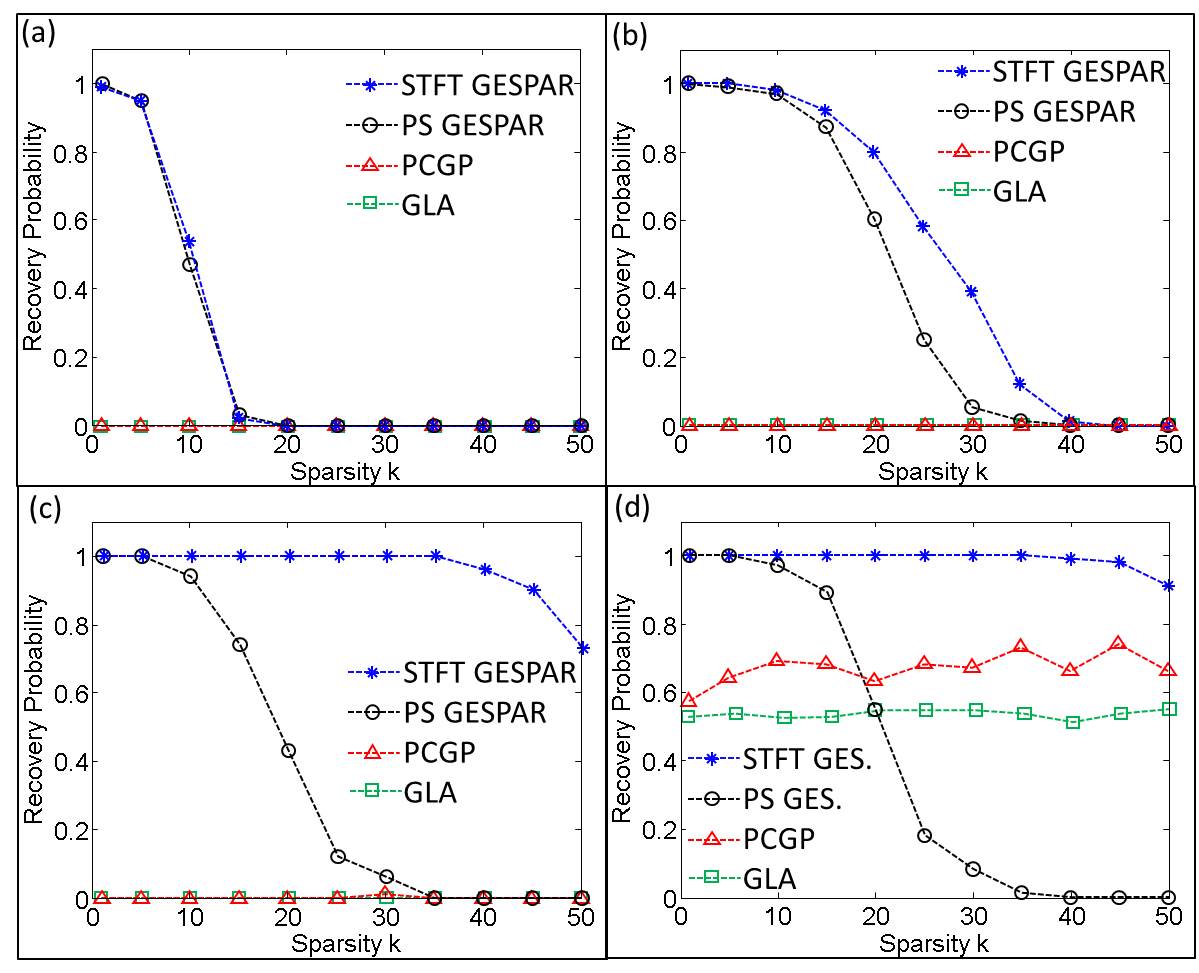}\label{fig:stftl}} \hspace{2.5cm}
\subfloat[{ MSE vs sparsity in the noisy setting for varying SNR (in [dB]) (i) 5, (ii) 15, (iii) 25, (iv) 35.}] {\includegraphics[scale=0.295]{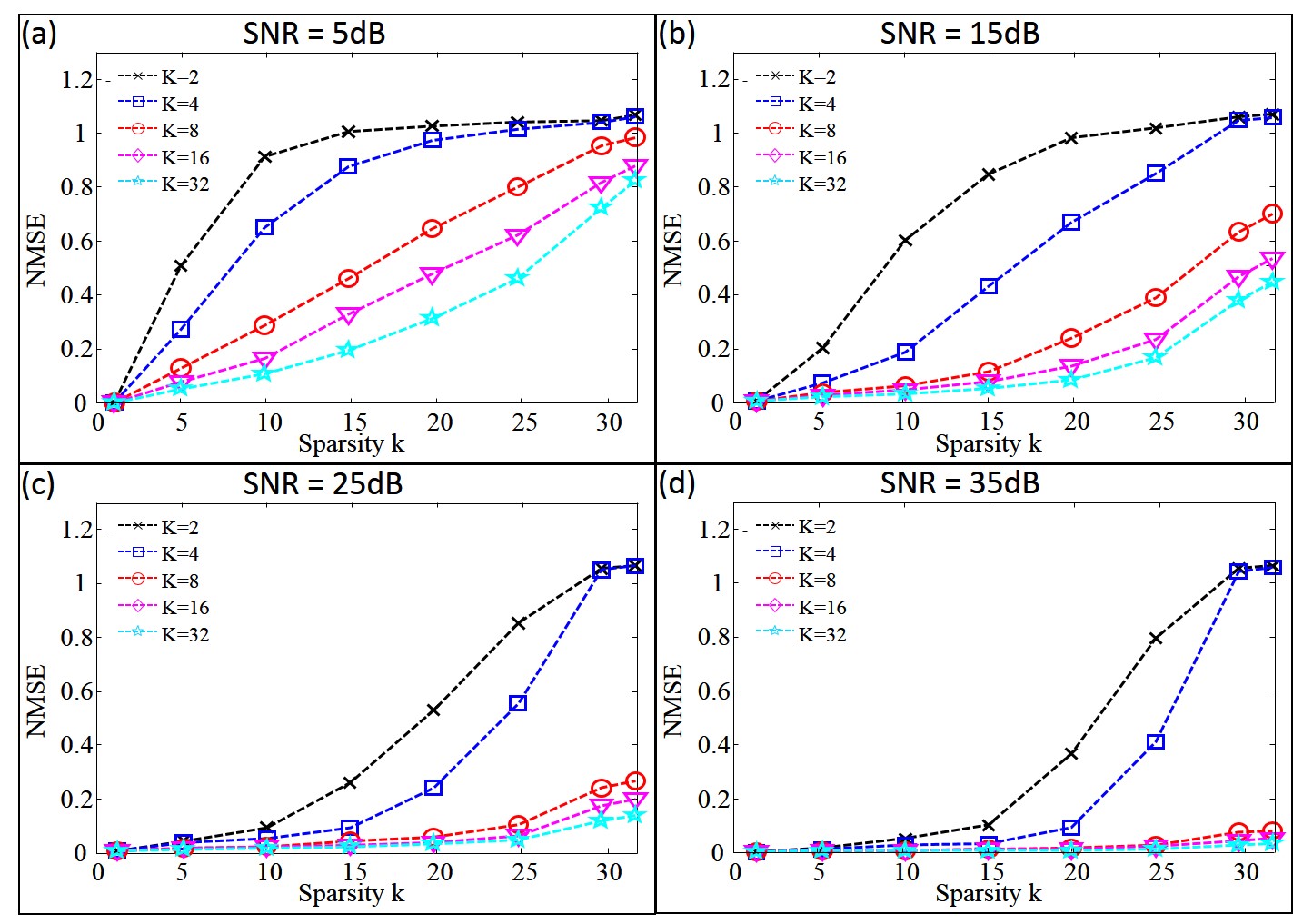}\label{fig:stftk}}
\end{center}
\caption{Performance of STFT-GESPAR for $N=\{32,64\}$ and various choices of $\{K,W,L\}$ (courtesy of \cite{eldarstft}).}
\end{figure}

STFT-GESPAR is used with a threshold $\tau=10^{-4}$ and maximum number of swaps $50000$, PCGP and GL are run using $50$ random initial points with $1000$ maximal iterations. For comparison with the Fourier transform approach, the performance of GESPAR with the same parameters and oversampling factors of $\{8,4,2,1\}$ is shown. In Figure \ref{fig:stftl}, the probability of successful recovery as a function of sparsity is plotted. Note that when $L=16$ there is no redundancy in the STFT and therefore it is not surprising that there is no advantage to the STFT method. In all other cases for which $L<16$, the STFT introduces redundancy, which leads to improved performance over simply oversampling the DFT. It is also evident that GESPAR outperforms both the GL and PCGP algorithms.

In Figure \ref{fig:stftk}, the effect of noise and the DFT length on the normalized mean-squared error  is considered. All parameters are the same as in Figure \ref{fig:stftl} besides $L$ which is set to $L=1$ and the signal length chosen as $N=32$. The DFT length is set to $K=\{2,4,8,16,32\}$. When $K$ is smaller than the window length $W=16$, a DFT of length $W$ is used and only the first $K$ measurements are chosen (i.e., only the $K$ low frequency measurements are used). As expected, increasing the DFT length improves the recovery ability. It is also evident that the performance improves significantly when all Fourier components are measured, namely, when $K \geq 16$.

%% file: Book_Chapter.bbl
\begin{thebibliography}{14}\small{

\bibitem{patt1} A. L. Patterson, ``A Fourier series method for the determination of the components of interatomic distances in crystals," Physical Review 46, no. 5 (1934): 372.

\bibitem{patt2} A. L. Patterson, ``Ambiguities in the X-ray analysis of crystal structures," Physical Review 65, no. 5-6 (1944): 195.

\bibitem{walther} A. Walther, ``The question of phase retrieval in optics," Journal of Modern Optics 10, no. 1 (1963): 41-49.

\bibitem{millane} R. P. Millane, ``Phase retrieval in crystallography and optics," JOSA A 7, no. 3 (1990): 394-411.

\bibitem{dainty} J. C. Dainty and J. R. Fienup, ``Phase retrieval and image reconstruction for astronomy,"  Image Recovery: Theory and Application (1987): 231-275.

\bibitem{rabiner} L. Rabiner and B. H. Juang, ``Fundamentals of speech recognition," Prentice Hall (1993).

\bibitem{stef} M. Stefik, ``Inferring DNA structures from segmentation data," Artificial Intelligence 11, no. 1 (1978): 85-114.

\bibitem{baykal} B. Baykal, ``Blind channel estimation via combining autocorrelation and blind phase estimation," IEEE Transactions on Circuits and Systems 51, no. 6 (2004): 1125-1131.

\bibitem{oppenheim1} A. V. Oppenheim and J. S. Lim, ``The importance of phase in signals," Proceedings of the IEEE 69, no. 5 (1981): 529-541.

\bibitem{eldarmagazine} Y. Shechtman, Y. C. Eldar, O. Cohen, H. N. Chapman, J. Miao and M. Segev, ``Phase retrieval with application to optical imaging," IEEE Signal Processing Magazine 32, no. 3 (2015): 87-109.
 
\bibitem{fienup0} J. R. Fienup, ``Phase retrieval algorithms: A personal tour [invited]," Applied Optics 52.1 (2013): 45-56.

\bibitem{hofstetter} E. M. Hofstetter, ``Construction of time-limited functions with specified autocorrelation functions," IEEE Transactions on Information Theory 10, no. 2 (1964): 119-126.

\bibitem{hayes} M. H. Hayes, ``The reconstruction of a multidimensional sequence from the phase or magnitude of its Fourier transform," IEEE Transactions on Acoustics, Speech and Signal Processing 30, no. 2 (1982): 140-154.

\bibitem{gerchberg} R. W. Gerchberg and W. O. Saxton, ``A practical algorithm for the determination of the phase from image and diffraction plane pictures," Optik 35 (1972): 237.

\bibitem{fienup} J. R. Fienup, ``Phase retrieval algorithms: A comparison," Applied Optics 21, no. 15 (1982): 2758-2769. 

\bibitem{bauschke} H. H. Bauschke, P. L. Combettes and D. R. Luke, ``Phase retrieval, error reduction algorithm, and Fienup variants: A view from convex optimization," JOSA A 19, no. 7 (2002): 1334-1345.

\bibitem{marchesini} S. Marchesini, ``Invited article: A unified evaluation of iterative projection algorithms for phase retrieval," Review of Scientific Instruments 78, no. 1 (2007): 011301.

{\edit \bibitem{yoninabook1} D. P. Palomar and Y. C. Eldar, ``Convex optimization in signal processing and communications," Cambridge University Press (2010).

\bibitem{yoninabook2} Y. C. Eldar and G. Kutyniok, ``Compressed sensing: Theory and applications," Cambridge University Press (2012).

\bibitem{yoninabook3} Y. C. Eldar, ``Sampling theory: Beyond bandlimited systems," Cambridge University Press (2015).

\bibitem{candesl0} E. J. Candes and T. Tao, ``Decoding by linear programming," IEEE Transactions on Information Theory 51, no. 12 (2005): 4203-4215.

\bibitem{candesmc} E. J. Candes and B. Recht, ``Exact matrix completion via convex optimization," Foundations of Computational Mathematics 9, no. 6 (2009): 717-772.

\bibitem{miao} J. Miao, P. Charalambous, J. Kirz and D. Sayre, ``Extending the methodology of X-ray crystallography to allow imaging of micrometre-sized non-crystalline specimens," Nature 400, no. 6742 (1999): 342-344.}

{\edit \bibitem{baraniukcpr} M. L. Moravec, J. K. Romberg and R. G. Baraniuk, ``Compressive phase retrieval,"  International Society for Optics and Photonics (2007): 670120-670120.}

\bibitem{candespr} E. J. Candes, Y. C. Eldar, T. Strohmer and V. Voroninski, ``Phase retrieval via matrix completion," SIAM Journal on Imaging Sciences 6, no.1 (2013): 199-225.

\bibitem{kishoreorig} K. Jaganathan, S. Oymak and B. Hassibi, ``Recovery of sparse 1-D signals from the magnitudes of their Fourier transform," IEEE International Symposium on Information Theory Proceedings  (2012): 1473-1477.

\bibitem{eldarcvx} Y. Shechtman, Y. C. Eldar, A. Szameit and M. Segev, ``Sparsity based sub-wavelength imaging with partially incoherent light via quadratic compressed sensing," Optics Express 19 (2011): 14807-14822.

\bibitem{kishorej} K. Jaganathan, S. Oymak and B. Hassibi, ``Sparse phase retrieval: Uniqueness guarantees and recovery algorithms," arXiv:1311.2745 (2015).

\bibitem{kishore2} K. Jaganathan, S. Oymak and B. Hassibi, ``Sparse phase retrieval: Convex algorithms and limitations," IEEE International Symposium on Information Theory Proceedings (2013): 1022-1026.
 
\bibitem{eldargespar} Y. Shechtman, A. Beck and Y. C. Eldar, ``GESPAR: Efficient phase retrieval of sparse signals," IEEE Transactions on Signal Processing 62, no. 4 (2014): 928-938.

\bibitem{vetterli}Y. M. Lu and M. Vetterli, ``Sparse spectral factorization: Unicity and reconstruction algorithms," IEEE International Conference on Acoustics, Speech and Signal Processing (2011): 5976-5979.

\bibitem{mukherjee} S. Mukherjee and C. Seelamantula, 	``An iterative algorithm for phase retrieval with sparsity constraints: Application to frequency domain optical coherence tomography," IEEE International Conference on Acoustics, Speech and Signal Processing (2012): 553Ð556.

\bibitem{ranieri} J. Ranieri, A. Chebira, Y. M. Lu and M. Vetterli, ``Phase retrieval for sparse signals: Uniqueness conditions,"  arXiv:1308.3058 (2013).

\bibitem{ohlsson2} H. Ohlsson and Y. C. Eldar, ``On conditions for uniqueness in sparse phase retrieval," IEEE International Conference on Acoustics, Speech and Signal Processing (2014): 1841-1845.

\bibitem{pfeiffer} I. Johnson, K. Jefimovs, O. Bunk, C. David, M. Dierolf, J. Gray, D. Renker and F. Pfeiffer, ``Coherent diffractive imaging using phase front modifications," Physical Review Letters 100, no. 15 (2008): 155503.

\bibitem{liu} Y. J. Liu et al. ``Phase retrieval in X-ray imaging based on using structured illumination," Physical Review A 78, no. 2 (2008): 023817.

\bibitem{popov} E. G. Loewen and E. Popov, ``Diffraction gratings and applications," CRC Press (1997).

\bibitem{faridian} A. Faridian, D. Hopp, G. Pedrini, U. Eigenthaler, M. Hirscher, and W. Osten, ``Nanoscale imaging using deep ultraviolet digital holographic microscopy," Optics Express 18, no. 13 (2010): 14159-14164.

\bibitem{trebino} R. Trebino, ``Frequency-resolved optical gating: The measurement of ultrashort laser pulses," Springer (2002).

\bibitem{rodenburg} J. M. Rodenburg, ``Ptychography and related diffractive imaging methods,"  Advances in Imaging and Electron Physics 150 (2008): 87-184.

{\edit \bibitem{kishorestft} K. Jaganathan, Y. C. Eldar and B. Hassibi, ``STFT phase retrieval: Uniqueness guarantees and recovery algorithms," arXiv:1508.02820 (2015).}

\bibitem{humphry} M. J. Humphry, B. Kraus, A. C. Hurst, A. M. Maiden, J. M. Rodenburg, ``Ptychographic electron microscopy using high-angle dark-field scattering for sub-nanometre resolution imaging," Nature Communications 3 (2012): 730.

\bibitem{yang} G. Zheng, R. Horstmeyer and C. Yang, ``Wide-field, high-resolution Fourier ptychographic microscopy," Nature photonics 7, no. 9 (2013): 739-745.

\bibitem{eldarstft} Y. C. Eldar, P. Sidorenko, D. G. Mixon, S. Barel and O. Cohen, ``Sparse phase retrieval from short-time Fourier measurements," IEEE Signal Processing Letters 22, no. 5 (2015): 638-642.

\bibitem{fazel1} B. Recht, M. Fazel and P. Parrilo, ``Guaranteed minimum-rank solutions of linear matrix equations via nuclear norm minimization," SIAM Review 52, no. 3 (2010): 471-501.

\bibitem{beck} A. Beck and Y. C. Eldar, ``Sparsity constrained nonlinear optimization: Optimality conditions and algorithms," SIAM Journal on Optimization 23, no. 3 (2013): 1480-1509.

{\edit \bibitem{yoninapr} Y. C. Eldar and S. Mendelson, ``Phase retrieval: Stability and recovery guarantees,"  Applied and Computational Harmonic Analysis 36, no. 3 (2014): 473-494.

\bibitem{balan1} R. Balan, P. Casazza and D. Edidin, ``On signal reconstruction without phase," Applied and Computational Harmonic Analysis 20, no. 3 (2006): 345-356.

\bibitem{balan2} R. Balan, B. G. Bodmann, P. G. Casazza and D. Edidin, ``Painless reconstruction from magnitudes of frame coefficients," Journal of Fourier Analysis and Applications 15, no. 4 (2009): 488-501.

\bibitem{ohlsson} H. Ohlsson, A. Yang, R. Dong, and S. Sastry, ``Compressive phase retrieval from squared output measurements via semidefinite programming," arXiv:1111.6323 (2011).

\bibitem{afonso3} A. S. Bandeira, J. Cahill, D. G. Mixon and A. A. Nelson, ``Saving phase: Injectivity and stability for phase retrieval," Applied and Computational Harmonic Analysis 37, no. 1 (2014): 106-125.

\bibitem{li} X. Li and V. Voroninski, ``Sparse signal recovery from quadratic measurements via convex programming,"  SIAM Journal on Mathematical Analysis 45, no. 5 (2013): 3019-3033.

\bibitem{samet} S. Oymak, A. Jalali, M. Fazel, Y. C. Eldar and B. Hassibi, ``Simultaneously structured models with application to sparse and low-rank matrices,"  IEEE Transactions on Information Theory 61, no. 5 (2015): 2886-2908.

\bibitem{sujay} P. Netrapalli, P. Jain and S. Sanghavi, ``Phase retrieval using alternating minimization,"  Advances in Neural Information Processing Systems (2013): 2796-2804.}

\bibitem{candespl} E. J. Candes, T. Strohmer, and V. Voroninski, ``Phaselift: Exact and stable signal recovery from magnitude measurements via convex programming," Communications on Pure and Applied Mathematics 66, no. 8 (2013): 1241-1274.

\bibitem{goemans} M. X. Goemans and D. P. Williamson, ``Improved approximation algorithms for maximum cut and satisfiability problems using semidefinite programming,"  Journal of the ACM 42, no. 6 (1995): 1115-1145.

\bibitem{daspremont} I. Waldspurger, A. d'Aspremont and S. Mallat, ``Phase recovery, maxcut and complex semidefinite programming," Mathematical Programming 149, no. 1-2 (2015): 47-81.

\bibitem{mahdi} E. J. Candes, X. Li, and M. Soltanolkotabi, ``Phase retrieval from coded diffraction patterns", arXiv:1310.3240 (2013).

\bibitem{wirtinger} E. J. Candes, X. Li and M. Soltanolkotabi, ``Phase retrieval via Wirtinger flow: Theory and algorithms," IEEE Transactions on Information Theory 61, no. 4 (2015): 1985-2007.

\bibitem{kishoreprm} K. Jaganathan, Y. C. Eldar and B. Hassibi, ``Phase retrieval with masks using convex optimization," IEEE International Symposium on Information Theory Proceedings (2015): 1655-1659 .

\bibitem{gross} D. Gross, F. Krahmer, and R. Kueng, ``Improved recovery guarantees for phase retrieval from coded diffraction patterns", arXiv:1402.6286 (2014).

\bibitem{afonso2} A. S. Bandeira, Y. Chen and D. G. Mixon, ``Phase retrieval from power spectra of masked signals," Information and Inference (2014): iau002.

\bibitem{kannan2} R. Pedarsani, K. Lee and K. Ramchandran, ``PhaseCode: Fast and efficient compressive phase retrieval based on sparse-graph codes,"  Annual Allerton Conference in Communication, Control, and Computing (2014): 842-849.

{\edit \bibitem{blindcs} S. Gleichman and Y. C. Eldar, ``Blind compressed sensing," IEEE Transactions on Information Theory 57, no. 10 (2011): 6958-6975.

\bibitem{fresnel} Y. Rivenson, A. Stern and B. Javidi, ``Compressive fresnel holography," Journal of Display Technology 6, no. 10 (2010): 506-509.

\bibitem{gazit} S. Gazit, A. Szameit, Y. C. Eldar and M. Segev, ``Super-resolution and reconstruction of sparse sub-wavelength images," Optics Express 17, no. 26 (2009): 23920-23946.

\bibitem{yoninanature} A. Szameit, Y. Shechtman, E. Osherovich, E. Bullkich, P. Sidorenko, H. Dana, S. Steiner, E. B. Kley, S. Gazit, T. Cohen-Hyams, S. Shoham, M. Zibulevsky, I. Yavneh, Y. C. Eldar, O. Cohen and M. Segev, ``Sparsity-based single-shot subwavelength coherent diffractive imaging," Nature Materials, Supplementary Info (2012).

\bibitem{subwave} C. Luo, S. G. Johnson, J. D. Joannopoulos and J. B. Pendry, ``Subwavelength imaging in photonic crystals," Physical Review B 68, no. 4 (2003): 045115.

\bibitem{superres} E. A. Ash and G. Nicholls, ``Super-resolution aperture scanning microscope," Nature (1972): 510-512. 

\bibitem{yoav} Y. Shechtman, Y. C. Eldar, O. Cohen and M. Segev, ``Efficient coherent diffractive imaging for sparsely varying dynamics," Optics Express 21, no. 5 (2013): 6327-6338.

\bibitem{yoninanature2} P. Sidorenko, A. Fleischer, Y. Shechtman, Y. C. Eldar, M. Segev and O. Cohen, ``Sparsity-based super-resolved coherent diffraction imaging of one-dimensional objects," Nature Communications 6 (2015).

\bibitem{waveguide} Y. Shechtman, E. Small, Y. Lahini, M. Verbin, Y. C. Eldar, Y. Silberberg and M. Segev, ``Sparsity-based super-resolution and phase-retrieval in waveguide arrays," Optics Express 21, no. 20 (2013): 24015-24024.

\bibitem{ankyl} J. Miao, C. Chen, Y. Mao, L. S. Martin and H. C. Kapteyn, ``Potential and challenge of ankylography," arXiv:1112.4459 (2011).

\bibitem{mutzafi} M. Mutzafi, Y. Shechtman, Y. C. Eldar, O. Cohen and M. Segev, ``Sparsity-based ankylography for recovering 3D molecular structures from single-shot 2D scattered light intensity," Nature Communications 6 (2015).}

\bibitem{heino} T. Heinosaari, L. Mazzarella and M. M. Wolf, ``Quantum tomography under prior information," Communications in Mathematical Physics 318, no. 2 (2013): 355-374.

\bibitem{afonso} A. S. Bandeira and D. G. Mixon, ``Near-optimal phase retrieval of sparse vectors," SPIE Optical Engineering Applications (2013): 88581O-88581O.

\bibitem{afonso1} B. Alexeev, A. S. Bandeira, M. Fickus and D. G. Mixon, ``Phase retrieval with polarization," SIAM Journal on Imaging Sciences 7, no. 1 (2014): 35-66.

\bibitem{kannan1} S. Pawar and K. Ramchandran, ``Computing a $k$-sparse $n$-length discrete Fourier transform using at most $4k$ samples and $o (k \log k)$ complexity," IEEE International Symposium on Information Theory Proceedings (2013): 464-468.

\bibitem{schafer} L. R. Rabiner and R. W. Schafer, ``Digital processing of speech signals," Prentice Hall (1978).

\bibitem{nawab} S. H. Nawab, T. F. Quatieri, and J. S. Lim, ``Signal reconstruction from short-time Fourier transform magnitude," IEEE Transactions on Acoustics, Speech and Signal Processing 31, no. 4 (1983): 986-998.

\bibitem{oppenheim} J. S. Lim and A. V. Oppenheim, ``Enhancement and bandwidth compression of noisy speech," Proceedings of the IEEE 67, no. 12 (1979): 1586-1604.

\bibitem{lim} D. Griffin and J. S. Lim, ``Signal estimation from modified short-time Fourier transform," IEEE Transactions on Acoustics, Speech and Signal Processing 32, no. 2 (1984): 236-243.

\bibitem{sun} D. L. Sun and J. O. Smith, ``Estimating a signal from a magnitude spectrogram via convex optimization," arXiv:1209.2076 (2012).

\bibitem{kane} D. J. Kane, ``Principal components generalized projections: A review [invited]," JOSA B 25, no. 6 (2008): A120-A132.

\bibitem{fienupnew} M. Guizar-Sicairos and J. R. Fienup, ``Phase retrieval with transverse translation diversity: A nonlinear optimization approach," Optics Express 16, no. 10 (2008): 7264-7278.

\bibitem{skiena} S. S. Skiena, W. D. Smith and P. Lemke, ``Reconstructing sets from interpoint distances (extended abstract),"  Annual Symposium on Computational Geometry (1990): 332-339.

\bibitem{dakic} T. Dakic, ``On the turnpike problem," PhD Thesis, Simon Fraser University (2000).

\bibitem{kishoreturn} K. Jaganathan and B. Hassibi, ``Reconstruction of integers from pairwise distances," arXiv:1212.2386 (2012).

\bibitem{papa} C. H. Papadimitriou and K. Steiglitz, ``Combinatorial optimization: Algorithms and complexity," Courier Corporation (1998).

\bibitem{bertsekas} D. P. Bertsekas, ``Nonlinear programming," (1999).

%\bibitem{fannjiang} A. Fannjiang, ``Absolute uniqueness of phase retrieval with random illumination," Inverse Problems 28, no. 7 (2012): 075008.

%\bibitem{fazel2} M. Fazel, H. Hindi, and S. Boyd, ``Log-det heuristic for matrix rank minimization with applications to Hankel and Euclidean distance matrices," American Control Conference (2003): 2156-2162.

%\bibitem{candesw} E. J. Candes, M. B. Wakin and S. P. Boyd, ``Enhancing sparsity by reweighted $l_1$ minimization," Journal of Fourier Analysis and Applications 14, no. 5-6 (2008): 877-905.

}
\end{thebibliography}
